\documentclass[11pt,a4,njp]{iopart}
\usepackage{iopams} 
\usepackage{stackrel}
\usepackage{amstext}
\usepackage{amssymb}
\usepackage{graphicx}
\usepackage{dcolumn}
\usepackage{bbold,bm}
\usepackage{mathrsfs}
\usepackage{xcolor}
\usepackage{cite}

\newcommand\vex[1]{\mathbf{#1}}
\newcommand\gvex[1]{\boldsymbol{#1}}

\def\bra#1{\mathinner{\langle{#1}|}}
\def\ket#1{\mathinner{|{#1}\rangle}}
\def\inner#1#2{\mathinner{\langle{#1}|{#2}\rangle}}

\def\bbra#1{\mathinner{\langle\hspace{-0.75mm}\langle{#1}|}}
\def\kket#1{\mathinner{|{#1}\rangle\hspace{-0.75mm}\rangle}}
\def\iinner#1#2{\mathinner{\langle\hspace{-0.75mm}\langle{#1}|{#2}\rangle\hspace{-0.75mm}\rangle}}

\def\sgn{\mathrm{sgn}}

\def\im{\mathrm{Im}\,}

\def\dd{\mathrm{d}}
\def\dbar{\,{\mathchar'26\mkern-11mu \mathrm{d}}}

\def\Texp{\mathrm{T}\hspace{-1mm}\exp}

\def\trans{{\raisebox{-1pt}{{\scriptsize\textsf{T}}}}}
\def\nodag{^{\vphantom{\dagger}}}

\usepackage[T3,T1]{fontenc}
\DeclareSymbolFont{tipa}{T3}{cmr}{m}{n}
\DeclareMathAccent{\invbreve}{\mathalpha}{tipa}{16}

\usepackage{accents}
\newlength{\hhatheight}
\newcommand{\hhat}[1]{%
    \settoheight{\hhatheight}{\ensuremath{\hat{#1}}}%
    \addtolength{\hhatheight}{-0.35ex}%
    \hat{\vphantom{\rule{1pt}{\hhatheight}}%
    \smash{\hat{#1}}}
}

\usepackage[normalem]{ulem}

\def\note#1{#1} 

\makeatletter
\newcommand{\colormathbox}[3][\mathord]{%
  #1{%
    \setlength{\fboxsep}{1pt}%
    \mathpalette\color@mathbox{{#2}{#3}}%
  }%
}
\newcommand{\color@mathbox}[2]{%
  \color@@mathbox#1#2%
}
\newcommand{\color@@mathbox}[3]{%
  \colorbox{#2}{$#1\m@th#3$}%
}
\makeatother

\def\mnote#1{#1} 

\def\strike#1{%
} 

\def\text#1{{\mathrm{#1}}}

\AtBeginDocument{}
\usepackage{hyperref}

\begin{document}

\title{Floquet Perturbation Theory: Formalism and {Application to Low-Frequency Limit}}

\author{M Rodriguez-Vega$^{1,3}$, M Lentz$^2$, and B Seradjeh$^{1,3}$}
\address{$^1$Department of Physics, Indiana University, Bloomington, IN 47405 USA}
\address{$^2$Department of Physics, Syracuse University, Syracuse, NY 13244 USA}
\address{$^3$Max Planck Institute for the Physics of Complex Systems, N\"othnitzer Stra\ss e 38, Dresden 01187 Germany}
\ead{babaks@indiana.edu}

\begin{abstract}
{We develop a low-frequency perturbation theory in the extended Floquet Hilbert space of a periodically driven quantum systems, which puts the high- and low-frequency approximations to the Floquet theory on the same footing.} 
It captures adiabatic perturbation theories recently discussed in the literature as well as diabatic deviation due to Floquet resonances. For illustration, we apply our Floquet perturbation theory to a driven two-level system as in the Schwinger-Rabi and the Landau-Zener-St\"uckelberg-Majorana models. We reproduce
some known expressions for transition probabilities in a simple and systematic way and clarify and extend their regime of applicability. We then apply the theory to a periodically-driven system of fermions on the lattice and obtain the spectral properties and the low-frequency dynamics of the system. 
\end{abstract}

\maketitle

\section{Introduction}

Driving a system's parameters periodically in time leads to qualitatively new phenomena that are absent in equilibrium. Well-known examples of such phenomena in classical systems include parametric resonance and stability~\cite{Arn88a}. In quantum systems, a well-known consequence of periodic driving is the Rabi oscillation in a two-level system~\cite{CohDupGry98a}. More recently, the repertoire of such phenomena has been expanded to many-body quantum systems{~\cite{pranjal2017,gorg2018,remi2017}}, including the appearance of non-equilibrium topological phases~\cite{OkaAok09a,JiaKitAli11a,LinRefGal11a,RecZeuPlo13a,WanSteJar13a,CayDorSim13a,KunFerSer14a,CarDelFru15a,ZouLiu16a,KliStaLos16a,ThaLosKli17a,Eck17a,RodSer17a,RodFerSer18a} and, in the presence of interactions and/or disorder, many-body localized phases~\cite{LazDasMoe14b,AbaDe-Huv16a,MoeSon17a} that exhibit subharmonic oscillations, thus realizing a time crystal~\cite{KheLazMoe16a,ElsBauNay16a,ZhaHesKyp17a,ChoChoLan17a}. {Also, recent experimental advances have allowed the realization of driven optical lattices~\cite{reitter2017,lohse2016,chen2011,struck2014,flaschner2018}.}

Analytically, the appearance of these novel features is usually understood within a high-frequency approximation, e.g. the rotating-wave approximation, Floquet-Magnus expansion, and Brillouin-Wigner theory~\cite{CasOteRos01a,ManCha11a,EckAni15a,MikKitYas15a,MohSaxKun16b}. These approximations often break down as frequency is lowered below the typical energy scale of the static system, such as the bandwidth or an equilibrium insulating gap. Though in certain cases, other perturbative schemes, such as the Schrieffer-Wolff theory~\cite{BukKolPol15a}, provide valuable insight away from the high-frequency regime, understanding the low-frequency behavior of these novel phases remains challenging. In the opposite limit of vanishingly small frequency, one may expect the dynamics to be governed by adiabatic evolution. Perturbative methods to account for diabatic correction to this adiabatic evolution have been developed~\cite{Teu03a,RigOrtPon08a,RigOrt10a,SheAshNor10a,MukMohSen17a}. However, the connection between these methods and the Floquet theory used for higher frequencies is not clear.

In this paper, we develop a systematic perturbation theory based on the Floquet theorem within the extended Floquet Hilbert space furnishing the steady states of a periodically driven quantum system~\cite{Shi65a,Sam73a}. Our approach is general and works whenever an operator in the Floquet Hamiltonian describing the dynamics of the system in the extended Floquet Hilbert space can be taken to be small. Indeed, we show how this Floquet perturbation theory leads to perturbative expansion both in the high- and the low-frequency limits. In both cases, we reproduce previous results in a compact and efficient way and show how higher-order terms are worked out systematically. Moreover, using this formalism we expand the applicability of these results and show when deviations are expected. In the low-frequency limit, we clarify the deviations from adiabatic evolution near quasienergy resonances~\cite{RusSan17a} that lead to Rabi oscillations. Finally, using our Floquet perturbation theory, we study a system of non-interacting fermions moving on a driven one-dimensional lattice at low frequency~\cite{RodSer17a,ChePanWan18a}. We derive the Floquet spectrum and show when the low-frequency limit does and does not approach the adiabatic evolution. 

We note that in the low-frequency limit the periodicity assumed in the Floquet theory is not a real restriction for reproducing the results of the adiabatic perturbation theory for a general drive. Basically, in this limit one can think of any drive as one big cycle of a periodic drive and find the desired evolution at any time mid cycle. The additional periodic structure in the Floquet theory is important only when one wishes to study the Floquet spectra of an actual periodic drive. One may call this low-frequency Floquet perturbation theory the ``Floquet adiabatic perturbation theory;'' however, this term is already used in the literature~\cite{WanZhoGon15a,WeiBukDAl17a,drese1999,novicenko2017} to describe the evolution of a driven system when a parameter of the drive is slowly varied. To avoid confusion, we do not use this terminology.

The paper is organized as follows. 
In Section~\ref{sec:FPT}, the Floquet perturbation theory is developed within the extended Floquet Hilbert space and used to derive high- and low-frequency series expansions of the Floquet spectrum. Formal aspects of the theory are presented in~\ref{app:Floquet}. In Section~\ref{sec:app}, we illustrate the formalism by applying it to transition probabilities in a driven two-level system, described separately by the Rabi-Schwinger and the Landau-Zener-St\"uckelberg-Majorana models. In Section~\ref{sec:dFATP}, we develop the degenerate low-frequency Floquet perturbation theory and demonstrate its application near quasienergy degeneracies in the low-frequency regime of the Landau-Zener model as well as the driven Su-Schrieffer-Heeger model of non-interacting fermions moving on a one-dimensional lattice. We conclude with a summary and outlook in Section~\ref{sec:sum}. Some technical details of our calculations are given in~\ref{app:LZinFAPT}.

\section{Floquet Perturbation Theory}\label{sec:FPT}

\subsection{Floquet Theory and the Extended Floquet Hilbert Space}

Floquet theorem is the statement that the solution to a differential equation with periodic coefficients can
be written as a phase factor multiplied by a periodic function. A direct consequence of this statement in
the condensed matter setting is the Bloch theorem for the solution to the Schr\"odinger equation in the presence of
a spatially periodic potential due to a lattice. In our discussion, we reserve the Floquet theorem for a system with
parameters that are periodic in {time}, $t$. 
The details of the Floquet theory formalism are presented in~\ref{app:Floquet}; here, we provide a summary.

For a Hamiltonian $\hat H(t)=\hat H(t+T)$ with period $T=2\pi/\Omega$, Floquet theorem states that the time-dependent Shr\"odinger equation $i\frac{\dd}{\dd t}|\psi(t)\rangle = \hat H(t)|\psi(t)\rangle$ takes steady-state solutions of the form 
\begin{eqnarray}
|\psi_\alpha(t)\rangle = e^{-i\epsilon_\alpha t}|\phi_\alpha(t)\rangle,
\end{eqnarray}
where the \emph{quasienergy} $\epsilon_\alpha$ is a conserved quantity and the periodic \emph{Floquet mode} $|\phi_\alpha(t)\rangle = |\phi_\alpha(t+T)\rangle$ satisfies the Floquet Schr\"odinger equation
\begin{eqnarray}\label{eq:FScht}
\left[\hat H(t)-i\frac{\dd}{\dd t}\right]|\phi_\alpha(t)\rangle = \epsilon_\alpha|\phi_\alpha(t)\rangle.
\end{eqnarray}
The $|\phi_\alpha(t)\rangle$ form a time-dependent orthonormal basis for the Hilbert space $\mathscr{H}$ and can be viewed as the eigenstates of the 
\note{\emph{time-dependent Floquet Hamiltonian}} 
$\hat H(t) - i\frac{\dd}{\dd t}$ with time-independent eigenvalues belonging to the \emph{Floquet zone}, $\epsilon_\alpha \in [-\Omega/2,\Omega/2]$. 
Using Floquet theorem, the evolution operator
\begin{eqnarray}
\hat U(t,t_0) = \Texp\left[-i\int_{t_0}^t \hat H(s)\dd s \right],
\end{eqnarray}
with $t_0<t$ and $\Texp$ the time-ordered exponential, can be decomposed as
\begin{eqnarray}
\hat U(t,t_0) = \hat \Phi(t) e^{-i(t-t_0)\hat H_F} \hat \Phi(t_0)^\dagger,
\end{eqnarray}
where
\numparts
\begin{eqnarray}
e^{-it\hat H_F} 
	&= \sum_\alpha e^{-i \epsilon_\alpha t} \ket{\phi_\alpha(0)}\bra{\phi_\alpha(0)}, \\
\hat\Phi(t) 
	&= \sum_\alpha \ket{\phi_\alpha(t)} \bra{\phi_\alpha(0)},
\end{eqnarray}
\endnumparts
define, respectively, the \emph{Floquet Hamiltonian} $\hat H_F$ and the \emph{micromotion operator} $\hat \Phi(t)$. \note{Here, we set $\hat \Phi(0) = \hat I$. We could choose a different boundary condition by a change of basis to $\ket{\gamma_\alpha} = \hat\Gamma^\dagger\ket{\phi_\alpha(0)}$, where $\hat\Gamma$ is a unitary operator. In this basis, the Floquet Hamiltonian is $\hat\Gamma^\dagger\hat H_F\hat\Gamma$ and the micromotion operator $\hat\Phi_\Gamma(t)=\sum_\alpha\ket{\phi_\alpha(t)}\bra{\gamma_\alpha}=\hat\Phi(t)\hat\Gamma$, with $\hat\Phi_\Gamma(0) = \hat\Gamma$. This freedom can lead to different truncated Floquet perturbative expansions, if $\hat\Gamma$ depends on the perturbation parameter itself~\cite{ManCha11a,EckAni15a}. 
We shall see an example of this in Sec.~\ref{sec:highF}. The evolution operator is independent of this choice.}

The structure we have described above can be formalized in terms of an extended Floquet Hilbert space $\mathscr{F}=\mathscr{H}\otimes\mathscr{I}$, where the auxiliary space $\mathscr{I}$ is the space of bounded periodic function over $[0,T)$~\cite{Sam73a}. 
We denote the states in $\mathscr{H, I},$ and $\mathscr{F}$ respectively by $\ket{\cdot}, |\cdot),$ and $\kket{\cdot}$ and the operators acting on each respective space as $\hat O, \invbreve O,$ and $\hhat O$. The space $\mathscr{I}$ is spanned by a continuous orthonormal basis $\{|t)\}, 0\leq t<T$,
\begin{eqnarray}
(t'|t) 
	= T\delta(t-t'), \quad
\int_0^T |t)(t| \frac{\dd t}T 
	= \invbreve I,
\end{eqnarray}
where $\invbreve I$ is the identity operator in $\mathscr{I}$. The auxiliary space $\mathscr{I}$ is also spanned by the orthonormal Fourier basis 
\begin{eqnarray}
|n) = \int_0^T e^{-in\Omega t}|t)\frac{\dd t}T, \quad n\in\mathbb{Z},
\end{eqnarray}
satisfying
\begin{eqnarray}
(n|m) = \delta_{nm},\quad
\sum_{n\in\mathbb{Z}}|n)(n| = \invbreve I.
\end{eqnarray}
We note
\begin{eqnarray}
|t) = \sum_{n\in\mathbb{Z}} e^{in\Omega t} |n).
\end{eqnarray}

A \emph{loop} in $\mathscr{H}$ {is a one-parameter family of states $\ket{\phi(t)}$ that is cyclic, i.e.} $\ket{\phi(T)}=\ket{\phi(0)}$. {It} can be \emph{lifted} to a loop in $\mathscr{F}$ given by $\kket{\phi_t}:=\ket{\phi(t)}|t)$. Associated with any loop $\kket{\phi_t}\in\mathscr{F}$ is the \emph{center} of the loop,
\begin{eqnarray}
\kket{\overline\phi} \equiv \int_0^T \kket{\phi_t}\frac{\dd t}T.
\end{eqnarray}
We define the Fourier-integral and the time-derivative operators,
\numparts
\begin{eqnarray}
\hhat \mu_n 
	&= \hat I\otimes\int_0^T |t)e^{{+}in\Omega t}(t|\frac{\dd t}T, \\
\hhat Z_{\mnote{t}}
	&= \hat I\otimes\sum_{n\in\mathbb{Z}} |n) n\Omega (n|,
	\label{eq:z_operator}
\end{eqnarray}
\endnumparts
such that for a loop $\kket{\phi_t}\in\mathscr{F}$,
\begin{eqnarray}
\hhat\mu_n\kket{\overline\phi} = \int_0^T e^{{+}in\Omega t}\kket{\phi_t} \frac{\dd t}T \equiv \kket{\phi_n},
\end{eqnarray}
is the $n$-th Fourier integral and
\begin{eqnarray}
\hhat Z_{\mnote{t}} \kket{\overline\phi} = \int_0^T i\frac{\dd \ket{\phi(t)}}{\dd t}|t) \frac{\dd t}T \equiv i\kket{\overline{{\dd\phi}/{\dd t}}},
\end{eqnarray}
is the center of the time-derivative of the loop. Then, the Floquet Schr\"odinger Eq.~(\ref{eq:FScht}) can be written in $\mathscr{F}$ as
\begin{eqnarray}\label{eq:FSch0}
(\hhat H - \hhat Z_{\mnote{t}}) \kket{\overline{\phi_\alpha}} = \epsilon_\alpha\kket{\overline{\phi_\alpha}},
\end{eqnarray}
{where $\hhat H = \int_0^T \hat H(t) \otimes |t)(t| \frac{\dd t}T$.}

Note, however, that the set $\{ \kket{\overline{\phi_\alpha}} \}$ of solutions to Eq.~(\ref{eq:FSch0}) is not large enough to furnish a complete basis for $\mathscr{F}$. Noting that $[\hhat H,\hhat\mu_n] = 0$ and
\begin{eqnarray}
{[ \hhat\mu_n, \hhat Z_{\mnote{t}}]} = n\Omega\hhat \mu_n,
\end{eqnarray}
we can write instead
\begin{eqnarray}\label{eq:FSch}
(\hhat H - \hhat Z_{\mnote{t}})\kket{\phi_{\alpha n}} = \epsilon_{\alpha n} \kket{\phi_{\alpha n}},
\end{eqnarray}
where $\epsilon_{\alpha n} \equiv \epsilon_\alpha{+}n\Omega$ and $\kket{\phi_{\alpha n}} \equiv \hhat \mu_n\kket{\overline{\phi_\alpha}}$. Indeed, $\hhat \mu_n$ is the ladder operator for $\hhat H - \hhat Z$, mapping the solution $\kket{\overline{\phi_\alpha}}$ with quasienergy $\epsilon_\alpha$ to $\kket{\phi_{\alpha n}}$ with quasienergy $\epsilon_\alpha {+} n\Omega$. 
Now, the solutions $\kket{\phi_{\alpha n}}$ to Eq.~(\ref{eq:FSch}) provide a full basis for $\mathscr{F}$.

\subsection{Floquet Perturbation Theory}
{Let us recap the Floquet perturbation theory~\cite{Shi65a} in the above language.}
The Floquet Schr\"odinger equation~(\ref{eq:FSch}) can be inverted in $\mathscr{F}$ to give the Floquet Green's function
\begin{eqnarray}\label{eq:FGreen}
\hhat G(\epsilon) \equiv (\epsilon-\hhat H+\hhat Z_{\mnote{t}})^{-1}=\sum_{\alpha n}\frac{\kket{\phi_{\alpha n}}\bbra{\phi_{\alpha n}}}{\epsilon-\epsilon_\alpha{-}n\Omega}.
\end{eqnarray}
The Floquet Green's function can be employed to calculate a variety of responses of the driven system. In this work, we focus on its application to perturbation theory. 

For the periodic Hamiltonian $\hat H(t)=\hat H_0(t)+\hat V(t)$, where $\hat H_0(t)$ is the unperturbed Hamiltonian and $\hat V(t)$ is the perturbing potential (both having a common period $T$), we lift $\hat H(t)$ to $\hhat H = \hhat H_0 + \hhat V$ in $\mathscr{F}$. Paralleling the conventional time-\emph{independent} perturbation expansion in $\mathscr{F}$, we then expand the solutions to the Floquet Schr\"odinger equation as
\numparts
\begin{eqnarray}
\epsilon_\alpha 
	&= \epsilon_{\alpha{(0)}} + \epsilon_{\alpha{(1)}} + \epsilon_{\alpha{(2)}} + \cdots, \\
\kket{\overline{\phi_\alpha}}
	&= \kket{\overline{\phi_{\alpha(0)}}} + \kket{\overline{\phi_{\alpha(1)}}} + \kket{\overline{\phi_{\alpha(2)}}} + \cdots,
\end{eqnarray}
\endnumparts
with $(\hhat H_0 - \hhat Z) \kket{{\overline{\phi_{\alpha(0)}}}} = \epsilon_{\alpha(0)} \kket{{\overline{\phi_{\alpha(0)}}}}$, to find for $i\geq 1$, 
\numparts
\begin{eqnarray}
\epsilon_{\alpha(i)}
	&= \bbra{\overline{\phi_{\alpha(0)}}}\hhat V\kket{\overline{\phi_{\alpha(i-1)}}}, \label{eq:FPTei}\\
\kket{\overline{\phi_{\alpha(i)}}}
	&= \hhat P_\alpha\hhat G_{0\alpha}\hhat P_\alpha\left[ \hhat V \kket{\overline{\phi_{\alpha(i-1)}}} - \sum_{j=1}^{i-1} \epsilon_{\alpha(i-j)} \kket{\overline{\phi_{\alpha(j)}}} \right], \label{eq:FPTpi}
\end{eqnarray}
\endnumparts
where $\hhat G_{0\alpha}=(\epsilon_{\alpha(0)}-\hhat H_0+\hhat Z_{\mnote{t}})^{-1}$ is the Floquet Green's function for $\hhat H_0$, $\hhat P_\alpha=\hhat I - \kket{\overline{\phi_{\alpha(0)}}}\bbra{\overline{\phi_{\alpha(0)}}}$ projects to the subspace of $\mathscr{F}$ that is orthogonal to $\kket{\overline{\phi_{\alpha(0)}}}$, and we have assumed the standard normalization $\iinner{\overline{\phi_{\alpha(0)}}}{\overline{\phi_\alpha}}=1$.

Explicitly,
\numparts
\begin{eqnarray}
\epsilon_{\alpha(1)} 
		&= \int_0^T \bra{\phi_{\alpha(0)}(t)}\hat V(t)\ket{\phi_{\alpha(0)}(t)} \frac{\dd t}T, \label{eq:FPTe1}\\
\kket{\overline{\phi_{\alpha(1)}}} 
	&= \sum_{(\beta, n)\neq(\alpha,0)}\frac{\bbra{\phi_{\beta n(0)}}\hhat V\kket{\overline{\phi_{\alpha(0)}}}}{\epsilon_{\alpha(0)}-\epsilon_{\beta(0)}{-}n\Omega}\kket{\phi_{\beta n(0)}}, \label{eq:FPTp1}
\end{eqnarray}
\endnumparts
and
\begin{eqnarray}
\epsilon_{\alpha(2)}
	&= \sum_{(\beta, n)\neq(\alpha,0)}\frac{|\bbra{\overline{\phi_{\alpha(0)}}}\hhat V\kket{\phi_{\beta n(0)}}|^2
	}{\epsilon_{\alpha(0)}-\epsilon_{\beta(0)}{-}n\Omega} \nonumber\\
	&= \sum_{{
	{\scriptsize \begin{array}{c} n\neq0 \\ \beta\neq\alpha \end{array}} }}\frac{\left\vert\int_0^T \bra{\phi_{\alpha(0)}(t)} e^{{+}in\Omega t}\hat V(t) \ket{\phi_{\beta(0)}(t)} \frac{\dd t}T\right\vert^2
	}{\epsilon_{\alpha(0)}-\epsilon_{\beta(0)}{-}n\Omega}, \label{eq:FPTe2}
\end{eqnarray}
etc.

\subsection{High-Frequency Expansion}\label{sec:highF}
As an example, we derive a high-frequency expansion using the Floquet perturbation theory {(see Ref.~\cite{EckAni15a} for a detailed discussion)}. Assuming the frequency is larger than the typical quasienergy, we shall take the unperturbed Hamiltonian $\hhat H_0=0$ and $\hhat V=\hhat H$. Therefore, $\epsilon_{\alpha(0)}=0$, and $\kket{\overline{\phi_{\alpha(0)}}}$ can be obtained from lifting an arbitrary time-independent set $\ket{\phi_\alpha}$ in $\mathscr{H}$ to $\mathscr{F}$. \note{Since the unperturbed quasienergies are degenerate for the same $n$, we employ degenerate perturbation theory, noting the matrix elements:
\begin{equation}
\bbra{\phi_{\alpha n}}\hhat{H}\kket{\phi_{\beta n}} = \bra{\phi_\alpha} \hat H^{(0)} \ket{\phi_\beta},
\end{equation}
where the Fourier components are $\hat H^{(n)} = \int_0^T e^{in\Omega t} \hat H(t) \frac{\dd t}T$. Thus, at the lowest order, $\ket{\phi_{\alpha}}$ are chosen as the eigenstates of $\hat H^{(0)}$.}
After some algebra{, using Eqns. (\ref{eq:FPTe1}), (\ref{eq:FPTp1}), and (\ref{eq:FPTe2}),} we find 
\numparts
\begin{eqnarray}
\epsilon_{\alpha(1)} 
	&= \bra{\phi_\alpha}\hat H^{(0)}\ket{\phi_\alpha},\\
\epsilon_{\alpha(2)}
	&= \bra{\phi_\alpha}\sum_{n\neq 0}\frac{[\hat H^{(-n)},\hat H^{\mnote{(n)}}]}{2n\Omega}\ket{\phi_\alpha},\\
\ket{\phi_{\alpha(1)}(t)} 
	&= \sum_{n\neq0}\frac{\hat H^{(\mnote{-}n)}e^{in\Omega t}}{n\Omega}\ket{\phi_\alpha}.\label{eq:highphi}
\end{eqnarray}
\endnumparts
\note{Thus, in the basis $\{\ket{\phi_\alpha}\}$,} the quasienergies are obtained by diagonalizing
\begin{eqnarray}
\hat H_{F}\equiv \hat H^{(0)}+\sum_{n\neq0}\frac{[\hat H^{(-n)},\hat H^{(n)}]}{2n\Omega}+O(1/\Omega^2),
\label{eq:hf_approx}
\end{eqnarray}
\note{and the micromotion takes the form
\begin{eqnarray}
\hat\Phi(t) 
	&= \sum_\alpha \ket{\phi_\alpha(t)}\bra{\phi_\alpha} \approx \hat I + \sum_{n\neq 0} \frac{e^{in\Omega t}}{n\Omega} \hat H^{(-n)}  \nonumber\\
	&\approx \exp\left[ i\sum_{n\neq 0} \frac{\hat H^{(-n)}e^{in\Omega t} - H^{(n)}e^{-in\Omega t}}{2in\Omega} \right].
\end{eqnarray}
}%
This is indeed the same expression obtained using other high-frequency expansions, such as 
\note{van-Vleck perturbation theory~\cite{ManCha11a,EckAni15a,MikKitYas15a}.
We note that in this basis, the boundary condition $\hat\Phi(0) = \hat I + \sum_{n\neq 0}\hat H^{(-n)}/(n\Omega) \neq \hat I$. Instead, $\int_0^T \log[\hat\Phi(t)] \frac{\dd t}{T} = 0$, again in agreement with the van-Vleck theory.}

\note{We can restore the boundary condition to identity by the unitary transformation $\hat\Phi(0)\ket{\phi_\alpha} = \ket{\phi_\alpha(0)}$ to the basis of \emph{perturbed} Floquet modes. In this basis, we obtain
\begin{eqnarray}
\hat H_F
	&\mapsto \hat \Phi(0) \hat H_F \hat \Phi^\dagger(0) \nonumber\\
	&\approx \hat H^{(0)} + \sum_{n\neq 0} \frac{[\hat H^{(-n)},\hat H^{(n)}]+[\hat H^{(0)},\hat H^{(n)}]+[\hat H^{(-n)},\hat H^{(0)}]}{2n\Omega},
\label{eq:FMHF}
\end{eqnarray}
and }
\begin{eqnarray}
\hat \Phi(t) 
	&= \sum_\alpha\ket{\phi_{\alpha}(t)}\bra{\phi_\alpha(0)}
	\approx \hat I + \sum_{n\neq 0} \frac{e^{in\Omega t} -1}{n\Omega} \hat H^{(\mnote{-}n)} \nonumber \\
	&\approx \exp\left[ i \sum_{n\neq 0}  
	\frac{e^{in\Omega t/2} \hat H^{(\mnote{-}n)} + e^{-in\Omega t/2} \hat H^{\mnote{(n)}}}{n\Omega} \sin\frac{n\Omega t}2\right].
\label{eq:FMPhi}
\end{eqnarray}
In the last step, we have written the micromotion in a form that is manifestly unitary. Note that \note{now} $\hat \Phi(0) \mnote{ = \sum_\alpha \ket{\phi_\alpha(0)}\bra{\phi_\alpha(0)} }= \hat I$ by \note{orthonormality of the Floquet modes. This boundary condition and Eqns.~(\ref{eq:FMHF}) and (\ref{eq:FMPhi}) agree with those obtained using the Floquet-Magnus expansion~\cite{ManCha11a,EckAni15a,MikKitYas15a}.}

\subsection{Low-Frequency Expansion}
A perturbative expansion at low frequencies can be obtained by rescaling time to $\tau = \Omega t$ and noting that the periodicity of the Hamiltonian $\hat H(\tau)$ is maintained when translating $\tau \to \tau + 2\pi$. The Floquet Schr\"odinger equation in rescaled units read
\begin{eqnarray}\label{eq:rFSch}
(\hhat H - \Omega \hhat Z_{\mnote{\tau}})\kket{\phi_{\alpha n}} = \epsilon_{\alpha n} \kket{\phi_{\alpha n}}{,}
\end{eqnarray}
{where \note{the dimensionless} $\hhat Z_{\mnote{\tau}}\mnote{=\hat{I}\otimes \sum_n |n) n (n|}$, and $\hhat H$ is defined below Eq. (\ref{eq:FSch0})}.
One may now attempt a perturbative expansion at low frequencies taking $-\Omega\hhat Z_{\mnote{\tau}}$ as the perturbation operator. However, there is a subtlety that must be addressed: the Floquet perturbation theory we developed in the previous section takes $\hhat Z_{\mnote{\tau}}$ as part of the unperturbed Hamiltonian. This is necessary to ensure that the eigenvalues of the unperturbed operator have the same modular structure as the final quasienergies{; that is, if $\epsilon$ is a quasienergy obtained from the perturbative solution of Eq.~(\ref{eq:rFSch}), then $\epsilon+n\Omega$ for any $n\in\mathbb{Z}$ should also be a quasienergy solution of Eq.~(\ref{eq:rFSch})}. By, taking $\hhat H$ as the unperturbed operator without including $\hhat Z_{\mnote{\tau}}$, the eigenvalues of the unperturbed Hamiltonian will no longer be modular. Indeed, the eigenstates of $\hhat H$ are nothing but the eigenstates of the instantaneous Hamiltonian $\hat H(\tau)$ lifted to $\mathscr{F}$:
\begin{eqnarray}\label{eq:FH0}
\hhat H \kket{\psi_{\alpha \tau}} = E_{\alpha \tau} \kket{\psi_{\alpha \tau}},
\end{eqnarray}
where $\kket{\psi_{\alpha \tau}}=\ket{\psi_\alpha(\tau)}|\tau)$, $\hat H(\tau)\ket{\psi_{\alpha}(\tau)} = E_{\alpha}(\tau)\ket{\psi_\alpha(\tau)}$, and $E_{\alpha \tau}=E_{\alpha}(\tau)$. The eigenvalues $E_{\alpha \tau}$ of $\hhat H$ are, therefore, not modular, unlike the eigenvalues $\epsilon_{\alpha n}$ of $\hhat H - \Omega \hhat Z$. {To avoid confusion, let us note that here $\tau$ is simply a label indexing the eigenvalues and eigenstates of $\hhat H$, even though the operator itself does not depend on a specific choice of this label.}

Therefore, in order to use perturbation theory to build the spectrum of $\hhat H - \Omega \hhat Z_{\mnote{\tau}}$ as a power series over the  spectrum of $\hhat H$, we need to amend our Floquet perturbation theory to ensure we obtain a modular spectrum. This can be done by using the general relationship{, employed in writing Eq.~(\ref{eq:FSch}),} between the {modular} Floquet spectrum and the {Fourier integrals} of {the loop} in $\mathscr{F}$ {obtained}  by lifting the loop of Floquet modes in $\mathscr{H}$. {Starting with the zeroth order solutions $\kket{\psi_{\alpha\tau}}$ in Eq.~(\ref{eq:FH0}), we} first use perturbation theory to find $\kket{\psi_{\alpha\tau(i)}}$ to the desired order $i$. The modular spectrum is then found by taking the Fourier transform of this loop in $\mathscr{F}$,
\begin{eqnarray}
\kket{\phi_{\alpha n(i)}} = \hhat\mu_n\kket{\overline{\psi_{\alpha(i)}}} = \int_0^{2\pi} e^{{+}in\tau}\kket{\psi_{\alpha \tau(i)}} \dbar\tau,
\end{eqnarray}
where $\dbar\tau\equiv \dd\tau/(2\pi)$. 
This defines the proper eigenstate of $\hhat H - \Omega\hhat Z_{\mnote{\tau}}$ with a modular eigenvalue $\epsilon_{\alpha n (i)} =  \epsilon_{\alpha (i)} {+} n\Omega$, and the quasienergy
\begin{eqnarray}
\epsilon_{\alpha (i)} = \bbra{\overline{\phi_{\alpha (0)}}}\hhat H\kket{\overline{\phi_{\alpha(i)}}} - \Omega \bbra{\overline{\phi_{\alpha (0)}}}\hhat Z\kket{\overline{\phi_{\alpha(i-1)}}}.
\end{eqnarray}
This equation follows from the Floquet Schr\"odinger equation and noting that
$
\iinner{\overline{\phi_{\alpha(0)}}}{\overline{\phi_{\alpha}}} = \int_0^{2\pi}\iinner{\psi_{\alpha\tau(0)}}{\psi_{\alpha\tau}}\dbar\tau =1.
$
Note that for $i\geq1$ the first term vanishes.
Explicitly, for the first few terms we find,
\numparts
\begin{eqnarray}
\epsilon_{\alpha(0)} 
	&= \int_0^{2\pi} E_{\alpha}(\tau) \dbar\tau; \label{eq:FAPTe0}\\
\epsilon_{\alpha(1)}
	&=\Omega\int_0^{2\pi} \bra{\psi_\alpha(\tau)}\frac1i\frac{\partial}{\partial\tau}\ket{\psi_\alpha(\tau)} \dbar\tau \label{eq:FAPTe1}; \\
\epsilon_{\alpha(2)}
	&= \Omega^2\int_0^{2\pi}\sum_{\beta\neq\alpha} \frac{\left\vert \bra{\psi_\beta} \frac1i\frac{\partial}{\partial\tau}\ket{\psi_\alpha} \right\vert^2}{E_\alpha(\tau)-E_\beta(\tau)} \dbar\tau \label{eq:FAPTe2};
\end{eqnarray}
\endnumparts
and
\begin{eqnarray}
\kket{\phi_{\alpha n(1)}} 
	= \Omega \int_0^{2\pi} \sum_{\beta\neq \alpha} \frac{\bra{\psi_\beta} \frac1i\frac{\partial}{\partial \tau}\ket{\psi_\alpha}}{E_\alpha(\tau)-E_\beta(\tau)}e^{{+}i n\tau}\kket{\psi_{\beta\tau}}\dbar\tau. \label{eq:FAPTp1}
\end{eqnarray}

There is one final loose end we now address: there is a gauge freedom in the choice of the instantaneous basis {$\ket{\psi_\alpha(\tau)}$} that we need to fix. Explicitly, $
\ket{\kappa_\alpha(\tau)} = e^{-i\Lambda_\alpha(\tau)} \ket{\psi_\alpha(\tau)}$ with $\Lambda_\alpha(2\pi)=\Lambda_\alpha(0)$ is another basis, satisfying
\begin{eqnarray}
[\hat H(\tau) - i\Omega\frac{\partial}{\partial\tau}]\ket{\kappa_\alpha(\tau)}
	&= [E_\alpha(\tau) - \Omega \frac{\partial \Lambda_\alpha}{\partial\tau}]\ket{\kappa_\alpha(\tau)} \nonumber \\
	&
	~~~- i\Omega e^{-i\Lambda_\alpha(\tau)}\frac{\partial}{\partial\tau}\ket{\psi_\alpha(\tau)}.
\end{eqnarray}
Thus, fixing the gauge by setting $\Omega\partial\Lambda_\alpha/\partial\tau = E_\alpha(\tau) -\epsilon_{\alpha(0)}$, that is,
\begin{eqnarray}
\Lambda_\alpha(\tau) =\frac1\Omega\int_0^\tau [E_\alpha(s)- \epsilon_{\alpha(0)}]\dd s,
\end{eqnarray}
up to a constant, we obtain
\begin{eqnarray}\label{eq:zeroF}
[\hat H(\tau) - i\Omega\frac{\partial}{\partial\tau}]\ket{\kappa_\alpha(\tau)}
	= \epsilon_{\alpha(0)}\ket{\kappa_\alpha(\tau)} + O(\Omega).
\end{eqnarray}
This is the zeroth-order {Floquet} Schr\"odinger equation.\footnote{{For completeness, we note that the gauge fixing can be done entirely  in $\mathscr{F}$ by defining the gauge transformation operator $\exp(-i\hhat\Lambda) = \int_0^{2\pi} e^{-i \Lambda_\alpha(\tau)} \kket{\psi_{\alpha \tau}}\bbra{\psi_{\alpha \tau}} \dbar \tau$. Then, $\kket{\kappa_{\alpha \tau}} = \exp(-i\hhat \Lambda)\kket{\psi_{\alpha \tau}}$, and Eq.~(\ref{eq:zeroF}) follows from the commutation relation $[\exp(-i\hhat\Lambda),\hhat Z_{\mnote{\tau}}] = \int_0^{2\pi} \frac{\dd\Lambda_\alpha(\tau)}{\dd\tau}\kket{\psi_{\alpha \tau}}\bbra{\psi_{\alpha \tau}} \dbar\tau$.}}
Therefore, we must indeed choose $\kket{\phi_{\alpha(0)}} = \kket{\overline{\kappa_\alpha}}$. This gauge-fixing was previously used by Martiskainen and Moiseyev~\cite{MarMoi15a}; however, they only justified its use numerically by showing that it improves the accuracy of the perturbative expansion. Here, we see that this gauge must be fixed for consistency of the adiabatic solution as the zeroth order term in the general low-frequency Floquet perturbation theory.

{Our low-frequency expansion is obtained for periodically driven systems using Floquet perturbation theory. However, this same approach can be used for general non-periodic and slowly driven systems by treating the whole evolution as one long single cycle of a periodic drive. In this way, we can connect our results to other low-frequency approximations, such as the adiabatic perturbation theory~\cite{Teu03a,RigOrtPon08a,RigOrt10a} and the adiabatic-impulse theory~\cite{SheAshNor10a,MukMohSen17a}. Our expressions obtained above for the evolution of the states are closely related to those obtained using the adiabatic perturbation theory~\cite{RigOrtPon08a,RigOrt10a}. Our method is, however, simpler in its structure and casts the entire procedure in the language of time-independent perturbation theory in the extended Floquet Hilbert space. The adiabatic-impulse theory~\cite{SheAshNor10a,MukMohSen17a}, on the other hand, requires the identification of special points during the drive where Landau-Zener transitions are likely to occur. These points are then treated separately from the rest of the drive, which is taken to be adiabatic. The accuracy of this approach depends strongly on the specific shape of the drive and lacks a natural low-frequency perturbation parameter. In contrast, our approach, similar to the adiabatic perturbation theory, can be used for any drive protocol. Finally, the low-frequency Floquet perturbation theory formulated in this work naturally connects to other approximate methods for higher frequencies that also use the structure of the extended Floquet Hilbert space.
}

\section{Applications}\label{sec:app}

\subsection{Schwinger-Rabi Model at Low Frequency}
As an example, take the matrix Hamiltonian~\cite{Sch37a}
\begin{eqnarray}
H(t) = \vex B(t) \cdot \gvex\sigma,
\end{eqnarray}
of a spin-$\frac12$ particle in a magnetic field $\vex B(t) $ rotating at frequency $\Omega$ and a fixed angle $\theta$ with the z-direction. (Here, $\gvex\sigma$ is the vector of Pauli matrices and  we set the magnetic moment to unity.) Taking $\vex n(t)=(\sin\theta\cos\Omega t,\sin\theta\sin\Omega t,\cos\theta)$ to be the unit vector in the direction of the magnetic field, and rescaling time as before to $\tau=\Omega t$, we have $H(\tau) = B \vex n(\tau) \cdot \gvex\sigma$.

The exact solution to the Schr\"odinger equation is found by going to the rotating frame given by the periodic unitary transformation $S(\tau) = e^{i\tau(1+\sigma_z)/2}$, where the Hamiltonian is
\begin{eqnarray}
H_\text{rot} 
	&= S(\tau)H(\tau)S^\dagger(t) - i \Omega S(\tau)\frac{\dd}{\dd \tau}S^\dagger(\tau) \nonumber\\
	&= -\frac{\Omega}2 + B \sin\theta\sigma_x + \left(B\cos\theta-\frac{\Omega}{2}\right)\sigma_z.
\end{eqnarray}
Since this is now time-independent, the solutions are found as the eigenstates $\ket{\chi_\pm}$ of $H_\text{rot}$ with eigenvalues $\epsilon_\pm = B e_\pm$, with $e_\pm = -(\Omega/2B) \pm \sqrt{(\cos\theta-\Omega/2B)^2+\sin^2\theta}$. In the original frame, we find the Floquet steady states $\ket{\phi_\pm(\tau)} = S^\dagger(\tau)\ket{\chi_\pm}$ with eigenvalues $\epsilon_\pm$ as quasienergies. To compare these exact solutions with the low-frequency Floquet perturbation theory, we impose the normalization $\inner{\chi_\pm(\Omega)}{\chi_\pm(\Omega=0)} = 1$; then, the Fourier transform of the loop $\ket{\phi_\pm(t)}$ lifted to $\mathscr{F}$ reads
\begin{eqnarray}
\kket{\phi_{\pm n}} 
	&= \left[\begin{array}{c} 
	(\cos\theta \pm e_\pm)\delta_{n,{+}1}
	\\
	\sin\theta\, \delta_{n,0}
	\end{array}\right] 
	\frac{|n)}{(1\pm e_\pm)f_\mp(\theta/2)},
\end{eqnarray}
with $f_+(x) = \sin x$ and $f_-(x) = \cos x$. Expanding in powers of $\Omega$, we find
\numparts
\begin{eqnarray}
\epsilon_{\pm(0)} &= \pm B; \label{eq:Rabie0}\\
\epsilon_{\pm(1)} &= -\Omega f_\mp^2(\theta/2); \label{eq:Rabie1}\\
\epsilon_{\pm(2)} &= \pm \frac{\Omega^2}{8B}\sin^2\theta,\quad \text{etc}; \label{eq:Rabie2}\\
\kket{\phi_{\pm n(0)}} &= \left[\begin{array}{c} 
\pm f_\mp(\theta/2) \delta_{n,{+}1} \\ f_\pm(\theta/2) \delta_{n,0} 
\end{array}\right] 
|n); \label{eq:Rabipsi0}\\
\kket{\phi_{\pm n(1)}} &= \frac{\Omega\sin\theta}{4B} \left[\begin{array}{c} 
- f_\pm(\theta/2) \delta_{n,{+}1} \\ \pm f_\mp(\theta/2) \delta_{n,0} \end{array}\right] 
|n),\quad \text{etc}. \label{eq:Rabipsi1}
\end{eqnarray}
\endnumparts

The low-frequency Floquet perturbation series is based on the instantaneous spectrum of $H(\tau)$ of spin Pauli matrices along $\vex n(\tau)$, given by the eigenstates
\begin{eqnarray*}
\ket{\psi_\pm(\tau)} = \left[\begin{array}{c} 
\pm f_\mp(\theta/2) e^{-i\tau} \\ f_\pm(\theta/2) \end{array}\right] 
\end{eqnarray*}
with eigenvalues $E_\pm(\tau) = \pm B$. Since the instantaneous eigenvalues are time-independent, the gauge $\Lambda_\pm=0$. 
Thus, at the lowest order, $\epsilon_{\pm(0)} = \int_0^{2\pi} E_\pm(\tau) \dbar\tau = \pm B$, and $\kket{\psi_{\pm n(0)}} = {\int_0^{2\pi}} e^{{+}i n\tau}\ket{\psi_\pm(\tau)}|\tau)\dbar\tau$, which reproduce Eqs.~(\ref{eq:Rabie0}) and~(\ref{eq:Rabipsi0}). Noting 
\numparts
\begin{eqnarray}
\bra{\psi_\pm(\tau)}\frac1i\frac{\partial}{\partial\tau}\ket{\psi_\pm(\tau)} 
	&= -f^2_\mp(\theta/2),\\
\bra{\psi_+(\tau)}\frac1i\frac{\partial}{\partial\tau}\ket{\psi_-(\tau)} 
	&= 
	\frac12\sin\theta,
\end{eqnarray}
\endnumparts
and using Eqs.~(\ref{eq:FAPTe1}),~(\ref{eq:FAPTp1}), and~(\ref{eq:FAPTe2}), we find precisely Eqs.~(\ref{eq:Rabie1}),~(\ref{eq:Rabipsi1}), and~(\ref{eq:Rabie2}).
{Thus, our low-frequency Floquet perturbation theory yields just the same leading order terms as those obtained from the exact solution.}

We shall now consider a more general periodic Hamiltonian $H(\tau)=\vex d(\tau)\cdot\gvex\sigma$, where $\vex d(\tau) = d(\tau)\vex n(\tau)$ is a vector, whose direction, $\vex n$, as well as its magnitude, $d$, change periodically in time. We denote the average $ \int_0^{2\pi} d(\tau)\dbar\tau \equiv \bar d$. Then, the Schr\"odinger equation is not, in general, exactly solvable. Transforming to the rotating frame, for example, will not produce a time-independent Hamiltonian any more. A high-frequency expansion can be developed in the rotating frame; however, this expansion fails at low enough frequency. Instead, we shall use the Floquet perturbation theory based on the instantaneous spectrum given by eigenstates $\ket{\psi_\pm(\tau)} = e^{\mp i\Lambda(\tau)}\left[\begin{array}{c} 
\pm f_\mp(\theta(\tau)/2) e^{-i\varphi(\tau)} \\ f_\pm(\theta(\tau)/2) \end{array}\right]$ 
and eigenvalues $E_\pm(\tau) = \pm d(\tau)$, where $\theta$ and $\varphi$ are the polar angles of $\vex n$ and the gauge $\Lambda(\tau) = \frac1\Omega\int_0^\tau [d(s) - \bar d]\dd s$. Now,
\numparts
\begin{eqnarray}
\bra{\psi_\pm}\frac1i\frac{\partial}{\partial\tau}\ket{\psi_\pm} 
	&= \frac{1\pm\cos\theta}2 \frac{\dd \varphi}{\dd \tau}, \\
\bra{\psi_-}\frac1i\frac{\partial}{\partial\tau}\ket{\psi_+} 
	&= \left( \frac{\sin\theta}2 \frac{\dd \varphi}{\dd \tau} - \frac i2 \frac{\dd \theta}{\dd \tau} \right) e^{-2i\Lambda}.
\end{eqnarray}
\endnumparts
So,
\numparts
\begin{eqnarray}
\epsilon_{\pm(0)} 
	&= \pm\int_0^{2\pi} d(\tau)\dbar\tau = \pm \bar d;\\
\epsilon_{\pm(1)} 
	&= \Omega \int_0^{2\pi} \frac{1\pm\cos\theta(\tau)}2 \frac{\dd\varphi}{\dd\tau} \dbar\tau; \\
\epsilon_{\pm(2)} 
	&= \pm\frac{\Omega^2}8 \int_0^{2\pi} \frac{(\frac{\dd\varphi}{\dd\tau})^2 \sin^2\theta(\tau)+(\frac{\dd\theta}{\dd\tau})^2 }{d(\tau)} {\dbar\tau}.
\end{eqnarray}
\endnumparts
The first-order correction to the Floquet steady state reads
\begin{eqnarray}
\ket{\psi_{\pm(1)}(\tau)} = \pm\frac{\Omega}2 \frac{\frac{\dd \varphi}{\dd \tau} \sin\theta \mp i \frac{\dd \theta}{\dd \tau}}{d(\tau)} e^{\mp 2i \Lambda(\tau)} \ket{\psi_\mp(\tau)}.
\end{eqnarray}

\subsection{Driven Landau-Zener Model}
Consider the Hamiltonian
\begin{eqnarray}\label{eq:dLZ}
H(\tau) = af(\tau)\sigma_x  + b\sigma_y,
\end{eqnarray}
where $f(\tau)$ is a periodic function satisfying $f(0)=-f(\pi)=1$ and $a, b>0$. Half way during the cycle, the first term switches sign, thus realizing the usual situation in the Landau-Zener model for large frequencies. 
In our general notation, $H(\tau )= d(\tau) \vex n(\tau)\cdot\gvex\sigma$, where $d(\tau) = \sqrt{a^2f^2(\tau)+b^2}$,  $\theta=\pi/2$, and $\cos\phi(\tau) = bf(\tau)/d(\tau)$. The minimum gap in the instantaneous spectrum is $b$ obtained when $f$ vanishes. We take $a\gg b$.

For $b=0$, the exact solution for the evolution operator is $U(\tau) = \exp[-i(a/\Omega)F(\tau)\sigma_x]$, where $F(\tau)=\int_0^\tau f(s)\dd s$. Thus, the Floquet spectrum is given by the quasienergies $\epsilon_{\pm(0)} = \pm aF(2\pi) /(2\pi)= \pm a f^{(0)}$ and Floquet steady states $\kket{\phi_{\pm \tau(0)}} = \frac1{\sqrt 2} 
[1 \: \pm1]^\trans
e^{\mp i (a/\Omega)\Lambda(\tau)}|\tau)$, with the micromotion phase $\Lambda(\tau) = F(\tau)-f^{(0)}\tau$. 

For $b/\Omega\ll 1$ or $b/a\ll 1$, we expect the operator $b\sigma_y$ to be ``small' compared to $\hhat H_0 - \Omega\hhat Z$ with $H_0(\tau) = af(\tau)\sigma_x$; thus, we may use the Floquet perturbation theory {in either the \emph{high-frequency} limit $\Omega\gg b$ or the \emph{high-amplitude} limit $a\gg b$} to find corrections to the Floquet spectrum:
\numparts
\begin{eqnarray}
\epsilon_{\pm(1)}
	&= 0, \\
\epsilon_{\pm(2)} 
	&= \pm \frac{b^2}\Omega \sum_{n}\frac{|g^{+}_n(2a/\Omega)|^2}{(2a/\Omega)f^{(0)} + n},\\
\kket{\overline{\phi_{\pm(1)}}} 
	&= \frac{b}\Omega\sum_{n,m}\frac{[g^{\pm}_n(2a/\Omega)]^*g^\pm_{n-m}(a/\Omega)}{(2a/\Omega)f^{(0)} \pm n} \frac 1{i\sqrt 2} \left[\begin{array}{c} 
	1 \\ \mp 1\end{array}\right] 
	|m),
\end{eqnarray}
\endnumparts
where $g_n^\pm(a/\Omega) = \int_0^{2\pi} e^{-in\tau}e^{\pm i (a/\Omega)\Lambda(\tau)}\dbar\tau$. Note that $[g_n^\pm(a/\Omega)]^* = g_{-n}^\mp(a/\Omega)$.
Thus, starting from an initial state $\ket{\phi_{\pm(0)}(0)}$, the probability of transitioning to state $\ket{\phi_{\mp(0)}(\tau)}$ at time $\tau$ is,
\begin{eqnarray}
P_\pm(\tau) 
	&= \left|\bra{\phi_{\mp(0)}(\tau)} U(\tau) \ket{\phi_{\pm(0)}(0)}\right|^2 \nonumber\\
	&= \left|\sum_\alpha\inner{\phi_{\mp(0)}(\tau)}{e^{-\frac i\Omega\epsilon_\alpha\tau}\phi_{\alpha}(\tau)}\inner{\phi_{\alpha}(0)}{\phi_{\pm(0)}(0)} \right|^2 \nonumber \\
	&\approx \left|\inner{\phi_{\mp(0)}(\tau)}{e^{-\frac i\Omega\epsilon_\pm\tau}\phi_{\pm(1)}(\tau)}\inner{\phi_{\pm(0)}(0)}{\phi_{\pm(0)}(0)} \right. \nonumber \\
	&~~~~+ \left. 
	\inner{\phi_{\mp(0)}(\tau)}{e^{-\frac i\Omega \epsilon_\mp\tau}\phi_{\mp(0)}(\tau)}\inner{\phi_{\mp(1)}(0)}{\phi_{\pm(0)}(0)}\right|^2, \nonumber \\
\end{eqnarray}
where the approximation is to the second-order in $b$. This yields, after some algebra,
\begin{eqnarray}
P_\pm(\tau) = \left| \frac b\Omega \int_0^\tau e^{\mp i (2a/\Omega) F(s)} \dd s \right|^2.
\end{eqnarray}

\begin{figure}[t]
\begin{center}
\includegraphics[width=4in]{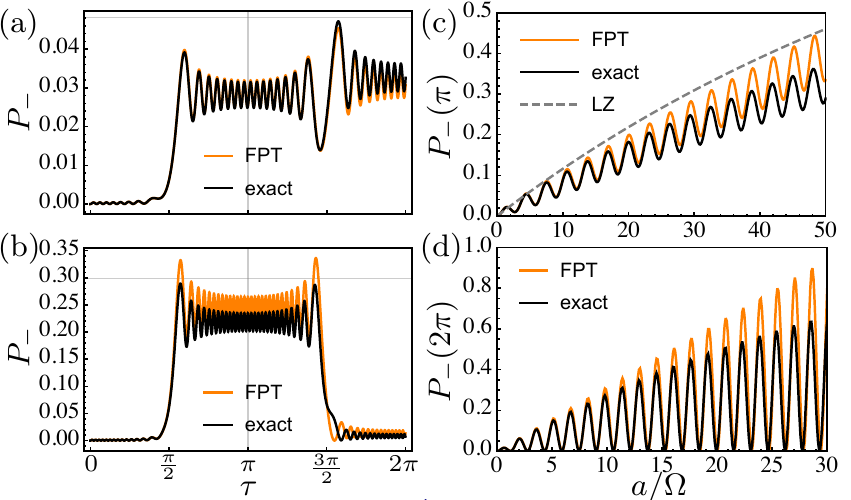}
\caption{The transition probability in the driven Landau-Zener model for the drive function $f(\tau) = \cos\tau$, calculated via Floquet perturbation theory (FPT) and numerically exactly, for (a) $b/\Omega=0.5$, $a/b=25$ and (b) $b/\Omega=1.8$, $a/b=45.1$. In (c) and (d) $a/b=20$. The Landau-Zener probability for the linear ramp at half cycle, Eq.~(\ref{eq:PLZ}), is shown by the horizontal grid line in (a) and (b) and by the dashed curve in (c).} \label{fig:LZcos}
\end{center}
\end{figure}

This expression can also be obtained~\cite{DavPec76a,Ber90a} directly from the Schr\"odinger equation written in the adiabatic basis, $\ket{\psi(\tau)} = \sum_{\alpha=\pm}A_\alpha(\tau)e^{-i\epsilon_{\pm(0)}\tau/\Omega}\ket{\phi_{\alpha(0)}(\tau)}$, which reads
\begin{eqnarray}
\frac{\dd A_\pm}{\dd\tau} = \pm \frac{b}{\Omega} e^{\pm i(2a/\Omega) F(\tau)} A_\mp(\tau).
\end{eqnarray}
Starting with initial state $\ket{\phi_{\pm(0)}(0)}$ and assuming $A_\pm(\tau) = 1 + O(b/\Omega)$ during the evolution, we find $A_\mp (\tau) \approx \mp (b/\Omega) \int_0^\tau e^{\mp i(2a/\Omega)F(s)} \dd s + O(b^2/\Omega^2)$. Our derivation using the more systematic Floquet perturbation theory, apart from being an application of the formalism, shows that this result is valid not only when $b/\Omega\ll 1$, but also when $b\ll a$, even for resonant $b\approx \Omega/2 \ll a$. In the latter case, the usual rotating-wave approximation for large frequencies fails; however, the highly oscillating phase factor suppresses higher order corrections. 

\begin{figure}[t!]
\begin{center}
\includegraphics[width=4in]{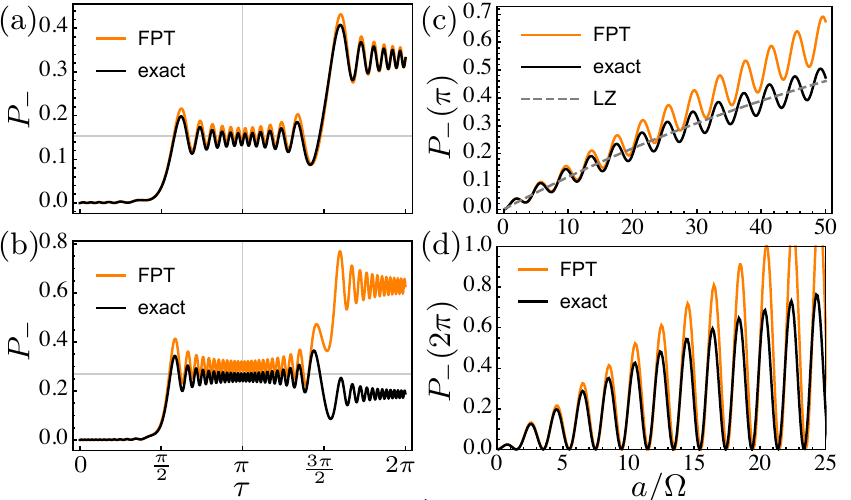}
\caption{The transition probability in the driven Landau-Zener model for the linear drive function, Eq.~(\ref{eq:lin}), calculated via Floquet perturbation theory (FPT) and numerically exactly, for (a) $b/\Omega=0.9$, $a/b=24$ and (b) $b/\Omega=1.8$, $a/b=51$. In (c) and (d) $a/b=20$. The Landau-Zener probability at half cycle, Eq.~(\ref{eq:PLZ}), is shown by the horizontal grid line in (a) and (b) and by the dashed curve in (c).} \label{fig:LZlin}
\end{center}
\end{figure}

In Figs.~\ref{fig:LZcos} and~\ref{fig:LZlin}, we compare $P_-(\tau)$ obtained by the exact numerical solution with the Floquet perturbation theory for $f(\tau)=\cos(\tau)$ as well as the sawtooth-linear function 
\begin{eqnarray}\label{eq:lin}
f(\tau) = 
\left\{ \begin{array}{lr}
\frac2\pi \left(\frac{\pi }{2}-\tau\right) & 0 \leq \tau \leq \pi, \\ 
\frac2\pi \left(\tau - \frac{3 \pi }{2} \right) & \pi \leq \tau \leq 2\pi. 
\end{array}\right.
\end{eqnarray}
Panels (a) and (b) in each figure show two choices of parameters at or near resonance $b/\Omega=0.5$. We note that while agreement is good in the first half of the cycle, it becomes less reliable in the second half. In fact, the end-of-cycle behavior of the Floquet perturbation theory is sensitive to the choice of parameters: for small final values of $P_-(\tau)$ around $\tau=2\pi$ the agreement is reasonably good, but for certain larger final values, as in Fig.~\ref{fig:LZlin}(b), the Floquet perturbation result becomes less reliable.

One way to understand these variations is to compare Floquet perturbation theory and exact results at half- and full-cycle. In panels (c) and (d) of each figure we compare our results for $\tau=\pi$ and $\tau=2\pi$, respectively, at a fixed ratio $b/a\ll 1$ as $a/\Omega$ is varied. For the cosine ramp, the Floquet perturbation theory gives,
$
P_-(\pi) = \pi ^2 b^2 \left([\mathbf{H}_0(2 a/\Omega)]^2+[J_0(2 a/\Omega)]^2\right),
$
and 
$
P_-(2\pi) = 4\pi^2(b/\Omega)^2 [J_0(2a/\Omega)]^2,
$
where $J_0$ and $\mathbf{H}_0$ are, respectively, Bessel and Struve functions.
For the linear ramp, we find
$
P_-(\pi) = [F_-^2(\sqrt{a/\Omega})+F_+^2(\sqrt{a/\Omega})]\pi ^2 b^2/(a\Omega)
$
where
$F_\pm(x) =\int _0^xf_\pm(\pi  y^2/2) dy$ are Fresnel integrals.
We note that for large $a/\Omega\gg 1$, the prefactor approaches $1/2$ and this reproduces the classic Landau-Zener-St\"uckelberg-Majorana formula~\cite{Lan32b,Zen32a,Stu32a,Maj32a},
\begin{eqnarray}\label{eq:PLZ}
P_{\text{LZ}} = 1 - \exp[-\pi^2 b^2/(2a\Omega)] \approx \pi^2 b^2/(2a\Omega).
\end{eqnarray}
However, the perturbative result starts to fail for larger $a/\Omega$ and fixed $b/\Omega$ since the magnitude of the exponent in $P_\text{LZ}$ becomes large. 
We show $P_{\text{LZ}}$ in panel (c) of each figure. We note that while the classic result
is in very good agreement with the exact result, it misses the oscillations as a function of $a/\Omega$. By contrast, the perturbative result captures these oscillations very well.

We also see the sporadic nature of the agreement between Floquet perturbation theory and the exact result in the second half-cycle. As seen in panels (d) of each figure, the final value $P_-(2\pi)$ shows large oscillations as a function of $a/\Omega$ and vanishes periodically. The Floquet perturbation theory result captures these oscillations and, in particular, the zeros of $P_-(2\pi)$ remarkably well. This feature is similar to the coherent destruction of transitions discussed in a periodically driven double-well potential~\cite{GroDitJun91a,GroJunDit91a,LloPla92a,LloPla94a}. For larger values of $a/\Omega$ going beyond its applicability, the Floquet perturbation theory overshoots the amplitude of the oscillations, eventually giving unphysical values larger than unity. This is due to the non-unitary nature of perturbative expansion of the micromotion operator. We expect that a more controlled expansion that respects the unitarity of the micromotion operator, similar to the Floquet-Magnus expansion, should resolve this problem.


\section{Degenerate Low-Frequency Floquet Perturbation Theory}\label{sec:dFATP}
In the preceding discussion we have assumed the quasienergy spectrum is non-degenerate. We shall now address the case where this is not true; as we will see this has interesting consequences for the dynamics. {Degenerate Floquet perturbation theory has been employeed before to investigate multiphoton excitations and heating processes in driven optical lattices. \cite{weinberg2015,strater2016}}

\subsection{Formalism}
First, let us say a few words about the different conventions for the range of quasienergies, which we formally restricted to the first Floquet zone $[-\Omega/2,\Omega/2]$. This choice is entirely arbitrary, of course, and it may be more suitable in some problems to make other choices. Shifting a quasienergy $\epsilon_{\alpha n}\mapsto\epsilon_{\alpha n} {+} m\Omega$ maps $\kket{\phi_{\alpha n}}\mapsto\hhat\mu_m\kket{\phi_{\alpha n}}=\kket{\phi_{\alpha n+m}}$. Doing so for all or just a subset of quasienergies has no physical effect. For example, in Eq.~(\ref{eq:FGreen}) all such shifts can be absorbed into a shift of the summation variable $n$. 
Now, in Floquet perturbation theory, even if the unperturbed quasienergies $\epsilon_{\alpha(0)}$ are in the first Floquet zone, the corrections, Eq.~(\ref{eq:FPTei}), may not be. Thus, in this section, we shall relax this condition.

Degeneracies arise when two or more quasienergies $\epsilon_{\alpha}^r$, labeled by $r$, coincide when shifted by integer multiples of frequency, $n^r$. For accidental or symmetry related degeneracies, $n^r=0$, and one might as well restrict the unperturbed quasienergies to the first Floquet zone from the outset. However, for low-frequency Floquet perturbation theory, where even the unperturbed quasienergies need to be calculated according to Eq.~(\ref{eq:FAPTe0}), it is more convenient to allow unperturbed quasienergies take values outside the first Floquet zone. In this case, we shall assume the shifts $n^r$ are unique. (Of course, one may always set one of them to zero.) {For initially \emph{non-degenerate} instantaneous energy eigenvalues $E_\alpha \neq E_\beta$, this happens only when the associated quasienergies become resonant at the Floquet zone center or edges; thus, $\epsilon_\alpha = \epsilon_\beta + n\Omega = m \Omega/2$ for some integer $m$.} 

This is an unusual situation in textbook perturbation theory, since the zeroth-order quantities are resonant up to first order in the small quantity $\Omega$. Nevertheless, we may still proceed by trying to find the proper superpositions of degenerate states that resolve the degeneracy. 
Define
\begin{eqnarray}
\kket{\chi_{\alpha}^r} = \sum_s c_\alpha^{rs} \hhat\mu_{n^r}\kket{\overline{\phi_{\alpha(0)}^s}},
\end{eqnarray}
with coefficients $c_\alpha^{rs}$ to be determined by solving
\begin{eqnarray}
(\hhat H - \Omega \hhat Z) \kket{\chi_{\alpha}^r} = (m\Omega/2 + \epsilon_{\alpha(1)}^r)\kket{\chi_{\alpha}^r} + O(\Omega^2).
\end{eqnarray}
This yields,
\begin{eqnarray}\label{eq:dFPTA1}
\sum_q W_\alpha^{rq} c_\alpha^{qs} = \epsilon_{\alpha(1)}^{s} c_\alpha^{rs},
\end{eqnarray}
where $W_\alpha$ is a matrix with diagonal elements $W_\alpha^{rr} = \epsilon_\alpha^r {+} n^r\Omega - m\Omega/2$, and the off-diagonal elements,
\begin{eqnarray}
W_\alpha^{r s} 
	&\equiv -\Omega \bbra{\overline{\phi^r_{\alpha}}} \hhat\mu_{n^r}^\dagger \hhat Z \hhat\mu_{n^s}\kket{\overline{\phi^s_{\alpha}}}, \quad r\neq s \nonumber \\
	&
	= \Omega \int_0^{2\pi} e^{i({n_s-n_r})\tau} \bra{\phi_\alpha^r(\tau)}\frac1i\frac{\partial}{\partial\tau}\ket{\phi_\alpha^s(\tau)} \dbar\tau, 
\end{eqnarray}
where we have used $\inner{\phi_\alpha^r(\tau)}{\phi_\alpha^s(\tau)} = 0$ for $r\neq s$.
The eigenvalue set of equations in~(\ref{eq:dFPTA1}) constitute the first-order degenerate low-frequency Floquet perturbation theory.

\subsection{Floquet Resonances in Landau-Zener Model}\label{sec:LZdFAPT}
Let us revisit the driven Landau-Zener model, Eq.~(\ref{eq:dLZ}), in the low-frequency Floquet perturbation theory. Here, we shall take $f(\tau)=\cos\tau$. The instantaneous eigenvalues are $E_\pm(\tau) = \pm\sqrt{a^2\cos^2\tau + b^2}$; so, the lowest-order quasienergies are
\numparts
\begin{eqnarray}
\epsilon_{\pm(0)}
	&=
\pm \frac{2b}\pi E\left(-a^2/b^2\right), \\
\epsilon_{\pm(2)} 
	&= \pm \frac b{4\pi}\left[E(-a^2/b^2)-K(-a^2/b^2)\right] (\Omega/b)^2,
\end{eqnarray}
\endnumparts
where $E(x)$ 
and $K(x)$ 
are complete elliptic integrals. 
(The first-order correction to quasienergy vanishes.) For $\Omega/b\ll1$, one can always find quasienergy degeneracies by tuning $a/b$ and sufficiently large shifts $n$. However, we note quasienergy degeneracies also occur for $\Omega/b\gtrsim1$, so the associated dynamics is not restricted to (very) low frequencies.

To the lowest order in $\Omega$, the adiabatic solutions are
\begin{eqnarray}\label{eq:LZad}
\kket{\overline{\phi_{\pm(0)}}} = \int \frac{e^{\mp i\Lambda(\tau)}}{\sqrt2}
\left[\begin{array}{c} 
e^{-i\varphi(\tau)} \\ \pm 1
\end{array}\right] 
|\tau) \dbar\tau,
\end{eqnarray}
where $\cot\varphi(\tau) = (a/b)\cos\tau$ and the gauge $\Lambda(\tau)=\frac1\Omega \int_0^\tau [E_+(s)-\epsilon_{+(0)}]ds$ can also be expressed in terms of an incomplete elliptical integral.
Near a quasienergy degeneracy, $\epsilon_+ = n\Omega/2 + \Delta$ and $\epsilon_- = - n\Omega/2 - \Delta${; thus,}
in the adiabatic basis {$\{ \kket{\overline{\phi_{+(0)}}}, \hhat\mu_n \kket{\overline{\phi_{-(0)}}} \}$, $m=n$, and}, 
\begin{eqnarray}
W = \left[\begin{array}{cc} 
\Delta & \Omega {z_n^*} \\ \Omega {z_n} & -\Delta
\end{array}\right], 
\end{eqnarray}
where 
$z_n = {-} \frac12 \int_0^{2\pi} (d\varphi/d\tau) e^{in\tau}e^{2i\Lambda(\tau)} \dbar\tau$
is nonzero only for odd $n$. We derive a closed-form expression for $z_n$ in~\ref{app:LZinFAPT}. Writing $z_n = -i \zeta_n |z_n|$ with $\zeta_n = -\sgn(\im z_n)$, we find the solutions
\numparts
\begin{eqnarray}
\kket{\chi^+} &= \cos\frac{\theta_n}2 \kket{\overline{\phi_{+(0)}}} +
i \zeta_n \sin \frac{\theta_n}2 {\hhat\mu_n} \kket{\overline{\phi_{-(0)}}}, \label{eq:2ldFATPa}\\
\kket{\chi^-} &= \sin \frac{\theta_n}2 \kket{\overline{\phi_{+(0)}}} -
i \zeta_n \cos \frac{\theta_n}2 {\hhat\mu_n}\kket{\overline{\phi_{-(0)}}}, \label{eq:2ldFATPb}
\end{eqnarray}
\endnumparts
with 
$0<\theta_n<\pi$ and $\tan \theta_n = \Omega |z_n|/\Delta$. The quasienergy  degeneracy is lifted to
\begin{eqnarray}
\epsilon^\pm = 
{+}n\Omega/2\pm\sqrt{\Delta^2+|z_n|^2\Omega^2}.
\end{eqnarray}
Exactly at the degeneracy,  $\epsilon^\pm = {+} n\Omega/2\pm\Omega|z_n|$, $\theta_n=\pi/2$, and $\kket{\chi^\pm} = \frac1{\sqrt2} (\hhat\mu_n\kket{\overline{\phi_{+(0)}}} \pm i \zeta_n \kket{\overline{\phi_{-(0)}}})$.

\begin{figure}[t]
\begin{center}
\includegraphics[width=4in]{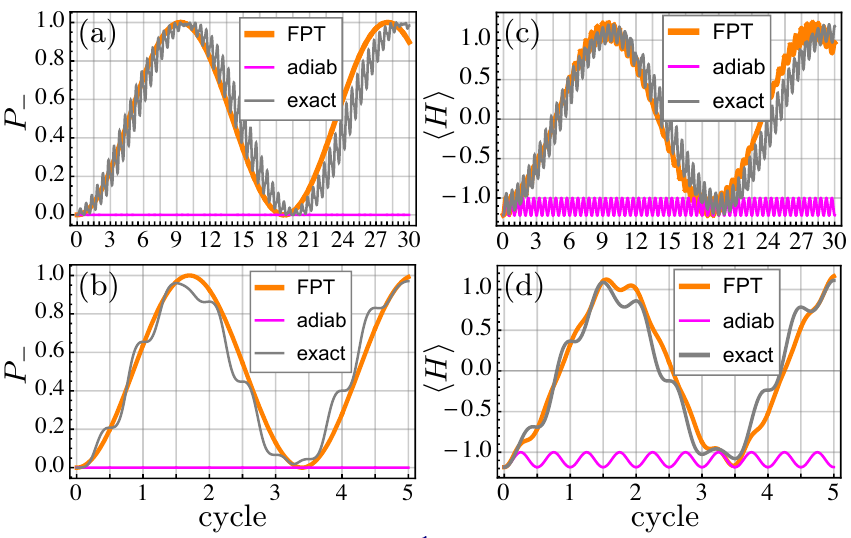}
\caption{The transition probability and energy vs. cycle time in the driven Landau-Zener model for the drive protocol $f(\tau) = \cos\tau$, starting from the ground state of initial Hamiltonian, calculated with degenerate low-frequency Floquet perturbation theory (FPT), the adiabatic approximation, and numerically exactly. The parameters $a/b = 0.691508, \Omega/b = 0.75$ (a, c) and $a/b = 0.631747, \Omega/b = 2.45$  (b, d), correspond to quasienergy degeneracies with, respectively, $n = 3$ and $n=1$.} \label{fig:FAPT}
\end{center}
\end{figure}

In Fig.~\ref{fig:FAPT}, we show the transition probability and the expectation value of energy at two such quasienergy degeneracies, one at a lower frequency $\Omega/b = 0.75$ and the other at a much higher frequency $\Omega/b=2.45$ with, respectively, $n=3$ and $n=1$. In both cases, we see that the low-frequency Floquet perturbation theory works well. This is remarkable for these frequencies are not particularly low compared to the minimum gap of $2b$. In fact, in the second case, the frequency $\Omega/b=2.45$ is larger than the gap, which is why it was chosen to obtain the lowest value of $n=1$. In this case, numerically calculated quasienergies are far from Floquet zone edges and simply eyeballing their evolution with $a/b$ does not hint at an avoided quasienergy degeneracy in the adiabatic approximation. Nevertheless, the lowest-order degenerate low-frequency Floquet perturbation theory produces a reasonably accurate result.

A few remarks are in order. The low frequency oscillations observed in Fig.~\ref{fig:FAPT} are manifestations of Rabi-like oscillations between resonant states $\kket{\overline{\phi_{+(0)}}}$ and ${\hhat\mu_n} \kket{\overline{\phi_{-(0)}}}$; at degeneracy, the frequency of these oscillations is $|z_n|\Omega$. These oscillations indicate a breakdown of adiabaticity~\cite{Ami09a,RusSan17a}: the system starting in the adiabatic state $\kket{\overline{\phi_{-(0)}}}$ would transition out fully over $1/(4|z_n|)$ cycles. 

However, one must be quite careful about statements of adiabatic breakdown at low frequencies. For smooth drive protocols, the splitting $|z_n|$ is typically exponentially small at low frequencies, meaning that a non-adiabatic transition would take exponentially long times. For example, in~\ref{app:LZinFAPT} we show that for $\Omega/b\ll 1$, a quasienergy degeneracy of the lowest shift order is obtained for $a\sim\Omega$ with $n\sim b\Omega$ and the quasienergy splitting vanishes as $(\Omega/b)^{2b/\Omega}$. Thus, the transition time out of the adiabatic evolution diverges as a factorially large number $\sim e^{-2(b/\Omega)\log(b/\Omega)}$. This trend is seen in Fig.~\ref{fig:FAPT}. The period of Rabi oscillations is multiplied by a factor of about six when going from a principal resonance at $n=1$ and $\Omega/b=2.45$ to the next resonance at $n=3$ and $\Omega=0.7$. Using low-frequency Floquet perturbation theory, we found that for the first resonance in the driven Landau-Zener model with $n=7$ and $\Omega/b=0.29$, this period rises up to more than $2\times10^5$ cycles.

We have shown here that the celebrated adiabatic approximation is the lowest order of the more general low-frequency Floquet perturbation theory, which is built on the Floquet Green's function. The latter correctly accounts for corrections to the adiabatic approximation and, in particular, its potential breakdown due to Rabi oscillations at quasienergy degeneracies. We must note, too, that not all quasienergy resonances at low frequency exhibit Rabi oscillations. The low-frequency Floquet perturbation theory shows when adiabatic evolution is preserved due to protected quasienergy crossings. For example, in our driven Landau-Zener model, $z_n=0$ for even $n$. 
At such crossings, Rabi oscillations become infinitely long, regardless of frequency, restoring adiabatic evolution in the degenerate subspace.

\subsection{Driven Su-Schrieffer-Heeger Model at Low Frequency}\label{sec:FTop}
As a second application, we apply the (degenerate) low-frequency Floquet perturbation theory to a driven lattice model of non-interacting fermions, namely the Su-Schrieffer-Heeger (SSH) model. Both the static~\cite{SuSchHee79a,MeiAnGad16a} and driven~\cite{RodSer17a,ChePanWan18a} versions of this model show distinct \emph{topological phases} that are distinguished by the appearance of protected bound states at the edges of the lattice with open boundary conditions or a topological winding number in the Brillouin zone for a system with periodic boundary conditions.

The Hamiltonian for the SSH model is written as
\begin{eqnarray}
\hat H = -\sum_{r=1}^{N-1} w_r \hat \xi_{r+1}^\dagger \hat\xi_r\nodag 
+ \hat H_b + \text{h.c.},
\end{eqnarray}
where $\hat \xi_r^\dagger$ is the creation operator of a (spinless) fermion at site $r$, 
$w_r=w+(-1)^r\delta$ is the modulated hopping amplitude, and $\hat H_b$ is a boundary Hamiltonian that depends on the choice of boundary conditions. In the following we take $w>0$ without loss of generality. 
For periodic boundary conditions, $\hat H_b = w_N \hat \xi^\dagger_N \hat\xi\nodag_1$, and even $N$ we label the two-point unit cells with $s=
\lfloor \frac{r+1}2\rfloor$, arrange the lattice operators into the spinor $\hat\Xi_s^\dagger=(\hat\xi^\dagger_{r\vphantom{+1}},\hat\xi^\dagger_{r+1})$, and write the mode expansion $\hat\Xi_s=\sum_{s}e^{iks}\hat\Xi_k$ with lattice momentum $k\in[-\pi,\pi]$ to find $\hat H=\sum_k \hat\Xi_k^\dagger h_k\nodag \hat\Xi_k\nodag$, with matrix Hamiltonian $h_k = \vex d_k\cdot\gvex\sigma$,
\begin{eqnarray}
d_{kx}+id_{ky} \equiv d_k = 2e^{ik/2}\left(w\cos\frac k2 + i \delta\sin\frac k2 \right),
\end{eqnarray}
and energies $E_{k\pm} = \pm2\sqrt{w^2\cos^2(k/2) + \delta^2\sin^2(k/2)}$.
We also note that the Hamiltonian can be mapped unitarily to $e^{i\frac k4\sigma_z}h_k e^{-i\frac k4\sigma_z} = (w\cos\frac k2) \sigma_x + (\delta\sin\frac k2) \sigma_y$.

\begin{figure}[t]
\begin{center}
\includegraphics[width=4in]{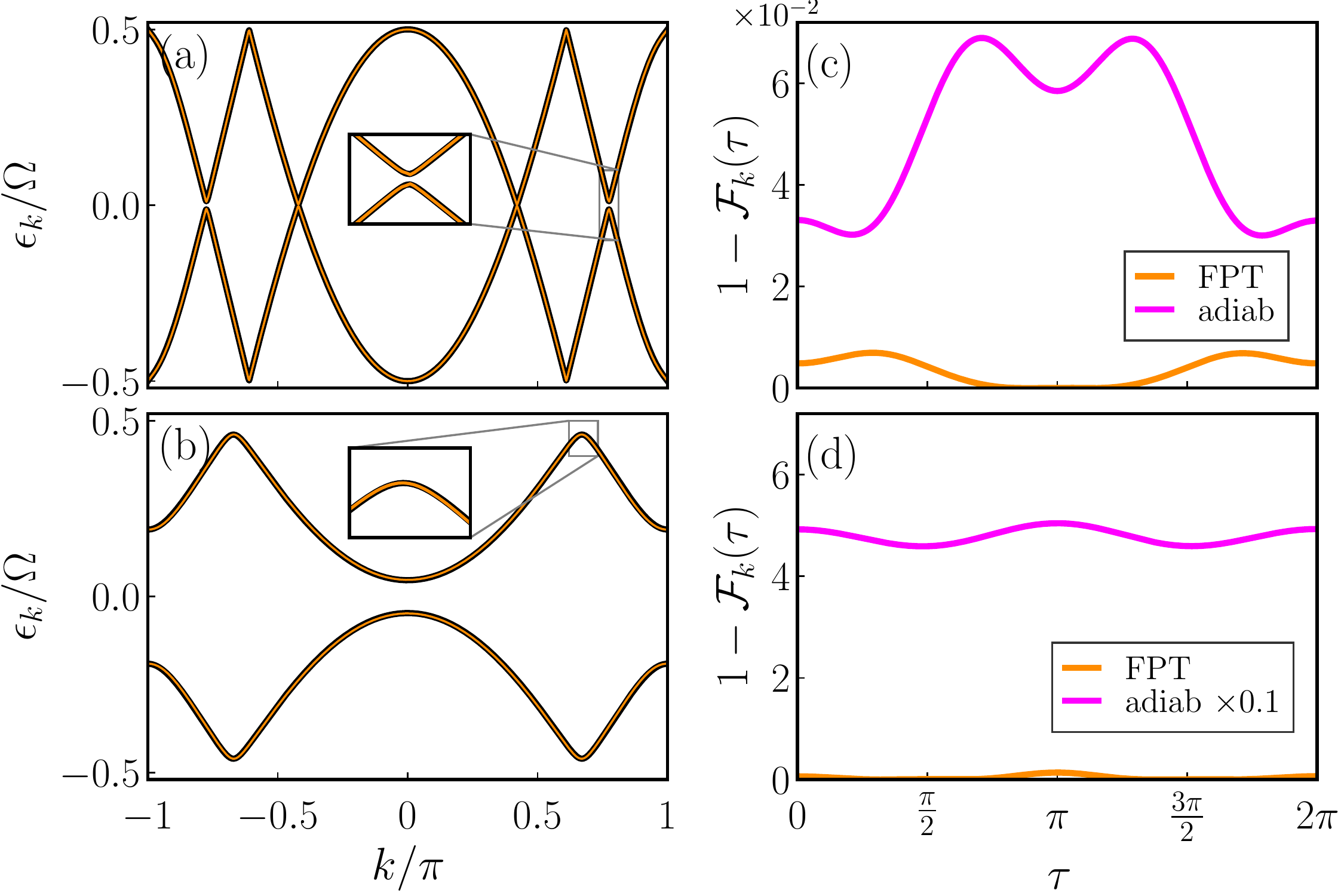}
\caption{Spectral properties of the driven Su-Schrieffer-Heeger model as a function of momentum $k$ for drive protocol $\delta(\tau) = \bar\delta + \delta_0 \cos\tau$. The quasienergy spectrum shown in (a) and (b) is found using exact numerical calculation (black) and the (degenerate) low-frequency Floquet perturbation theory (orange) for two different frequencies in (a) and (b). The insets show a closeup around quasienergy degeneracies. At these quasienergy degeneracies we show in (c) and (d) the infidelity in a single cycle, $1-\mathcal{F}$, where $\mathcal{F}$ is the modulus squared of the overlap between the numerically exact solution and the adiabatic solution (purple) or the lowest-order solution of the degenerate Floquet perturbation theory (orange).
The parameters are $\bar\delta/w=0.2$, $\delta_0/w = 0.1$, and $\Omega/w = 0.8$ (a, c) $\Omega/w = 2.1$ (b, d), and $k/\pi=0.767$ (c), $k/\pi=0.675$ (d).
}\label{fig:SSH}
\end{center}
\end{figure}

Now, we consider the driven SSH model, where $\delta(\tau)$ is a periodic function of time with frequency $\Omega$. We shall assume below that $|\delta| \leq w$ and denote the average $\bar\delta = \int_0^{2\pi} \delta(\tau) \dbar\tau$. In the low-frequency limit, to the lowest order in $\Omega$, the adiabatic solutions are
\begin{eqnarray}
\ket{\phi_{k\pm(0)}(\tau)} = \frac{e^{\mp i \Lambda_k(\tau) - i k \sigma_z/4}}{\sqrt{2}} 
	\left[\begin{array}{c} 
	e^{i \varphi_k(\tau)} \\
	\pm1
	\end{array}\right], 
\end{eqnarray}
with $\Lambda_k(\tau) = \frac{1}{\Omega} \int_0^\tau \left[ E_{k+}(s) - \epsilon_{k+(0)}\right]ds$ and $\tan \varphi_k(\tau) = -[\delta(\tau)/w] \tan\frac k2$.
The quasienergies in the leading-order non-degenerate low-frequency Floquet perturbation theory are $\epsilon_{k\pm}(\Omega) = \epsilon_{k\pm(0)} + \epsilon_{k\pm(1)} +  \epsilon_{k\pm(2)} + \cdots$, where
\numparts
\begin{eqnarray}
\epsilon_{k\pm(0)} 
	&= \int_{0}^{2\pi} E_{k\pm}(\tau) \dbar\tau, \\
\epsilon_{k\pm (1)} 
	&=  0, \\
\epsilon_{k\pm(2)} 
	&= \frac{\Omega^2}{8} \int_0^{2\pi} 
\frac{|{\partial \varphi_k}/{\partial \tau}|^2}{E_{k\pm}(\tau)} \dbar\tau.
\end{eqnarray}
\endnumparts

For a given $\Omega < 4w$ there are a set of points $\pm k^\wp(\Omega)$ in the Brillouin zone where quasienergies become degenerate: $\epsilon_{k^\wp {+} } = \epsilon_{k^\wp {-}} {+} n_{k^\wp} \Omega$, with $ \lceil 4|\bar\delta|/\Omega \rceil \leq n_{k^\wp} \leq \lfloor 4w/\Omega \rfloor$. Near these points we must employ degenerate low-frequency Floquet perturbation theory. Expand $\epsilon_{k\pm} = \pm(n_{k^\wp}\Omega/2 + \Delta_k^\wp)$ and $\kket{\chi_k^{\wp\pm}} = c_{k+}^\pm \kket{\phi_{k+(0)}} + c_{k-}^\pm {\hhat\mu_{n_{k^\wp}}} \kket{\phi_{k-(0)}}$, 
to find 
$W_k^\wp \left[\begin{array}{c} 
c_{k+}^\pm \\ c_{k-}^\pm  \end{array}\right] 
=  \epsilon_{k(1)}^{\wp\pm} \left[\begin{array}{c} 
c_{k+}^\pm \\ c_{k-}^\pm  \end{array}\right]$ 
with
$
W_{k}^\wp =  \left[\begin{array}{cc} 
	\Delta_k^\wp & \Omega {z^{\wp*}_{k}}  \\
	\Omega {z^{\wp}_{k}} & -\Delta_k^\wp
\end{array}\right] 
$.
Here, $z^\wp_k \equiv z_{k,n_{k^\wp}}$, where
\begin{eqnarray}
z_{k,n} = -\frac{1}{2} \int_0^{2\pi} e^{i n \tau} e^{2 i \Lambda_k(\tau )}\frac{\partial\varphi }{\partial \tau}\dbar\tau,
\end{eqnarray}
determines the gap opening at the degeneracy point as $\epsilon_{k(1)}^{\wp\pm} = \pm\sqrt{{\Delta_k^{\wp}}^2 + \Omega^2|z^\wp_k|^2}$ and the solutions $\kket{\chi_k^{\wp\pm}}$ in a fashion similar to Eqs.~(\ref{eq:2ldFATPa}) and~(\ref{eq:2ldFATPb}).

\begin{figure}[t]
\begin{center}
\includegraphics[width=4in]{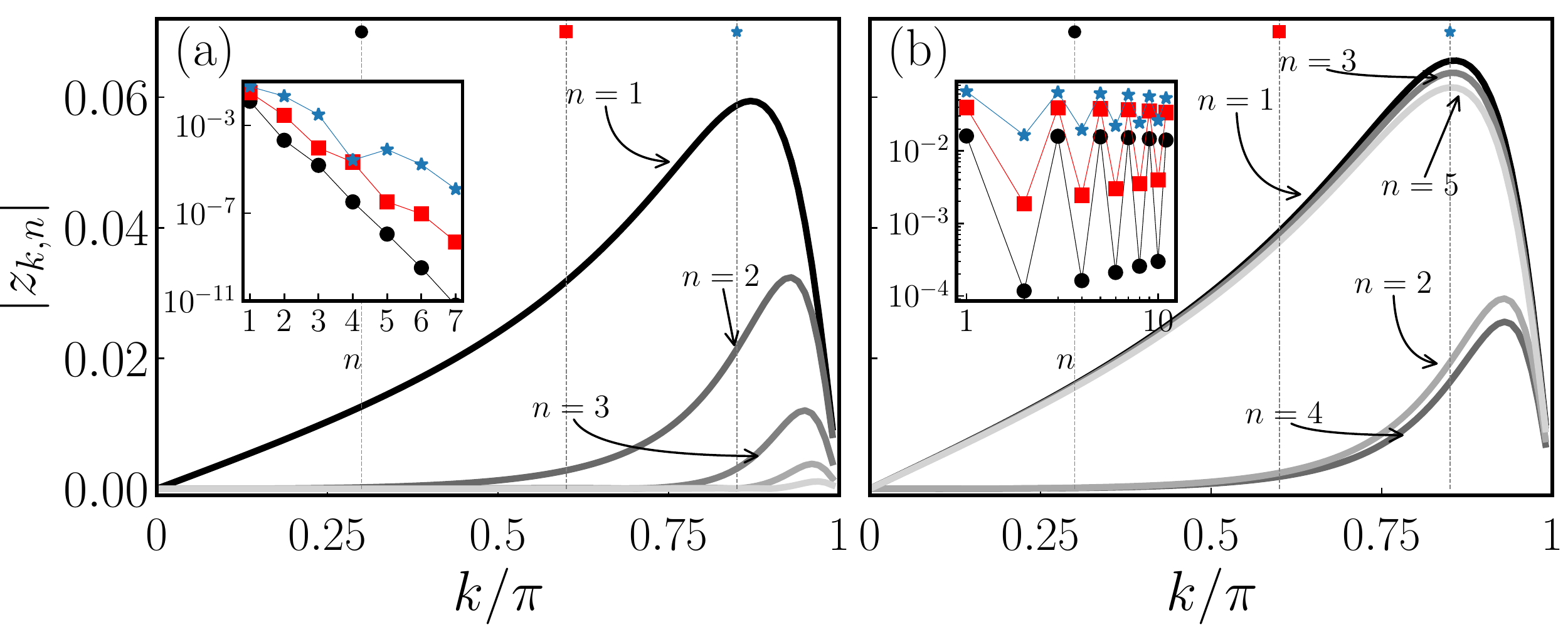}
\caption{
The off-diagonal element $|z_{k,n}|$ for a Floquet resonance of order $n$ in the driven Su-Schrieffer-Heeger model with the drive protocols $\delta(\tau)=\bar\delta + \delta_0 \cos \tau$ (a), and the smoothed two-step protocol $\delta(\tau)=\bar\delta + \delta_0 \arctan\left( B \sin \tau \right)$ (b). The values of parameters are: $\bar\delta = 0.2, \delta_0 = 0.1$, $\Omega/w=0.8$ and $B=20$. The insets show $|z_{k,n}|$ as a function of $n$ for the momenta indicated by the symbols.  
}\label{fig:SSHz}
\end{center}\vspace{-5mm}
\end{figure}

In Fig.~\ref{fig:SSH} we compare the spectral measures obtained from the adiabatic approximation and the low-frequency Floquet perturbation theory with the numerically exact solution for a smooth drive protocol $\delta(\tau) = \bar\delta + \delta_0 \cos\tau$. The quasienergies found using degenerate low-frequency Floquet perturbation theory match the exact solution remarkably well. The infidelity of an approximate solution, $\ket{\phi(\tau)}$, is defined as $1-\mathcal{F}$, where $\mathcal{F}(\tau) = |\inner{\phi_\text{e}(\tau)}{\phi(\tau)}|^2$ and $\ket{\phi_\text{e}(\tau)}$ is the exact solution. Near a quasienergy degeneracy point, the infidelity of the adiabatic approximation increases dramatically as the frequency increases. The infidelity of the solution obtained using the degenerate low-frequency Floquet perturbation theory, on the other hand, is not only small relative to the adiabatic approximation, but it also remains small on the absolute scale even for larger frequencies. This demonstrates the consistency and accuracy of the Floquet perturbation theory.

As the frequency is lowered, the order $n_{k^\wp}$ increases as $\sim 1/\Omega$. For a smooth drive protocol, the adiabatic limit is obtained as the off-diagonal element $z_k^\wp$ vanishes. For example, for the sinusoidal drive protocol, $z^\wp_k$ vanishes exponentially similar to the Floquet resonances in the Landau-Zener model. However, for a drive protocol with sharp features, such as a step-wise protocol, the approach to the adiabatic limit may be slower or even violated. In Fig.~\ref{fig:SSHz} we show the dependence of $z_{k,n}$ on $k$ and the order $n$ of the quasienergy degeneracy. Indeed, as shown in Fig.~\ref{fig:SSHz}(a) and the inset, for the sinusoidal drive protocol, $z_{k,n}$ vanishes with $n$ in an exponential manner. However, for step-wise protocol, shown in Fig.~\ref{fig:SSHz}(b), $z_{k,n}$ approaches a limiting value as $n$ increases that depends on the parity of $n$.

\section{Summary and Outlook}\label{sec:sum}

Floquet perturbation theory recasts time-dependent perturbation theory of a periodically driven quantum system in terms of a time-independent perturbation theory in the extended Floquet Hilbert space of the periodic operators. This formalism is transparent and strips away cumbersome book-keeping that is required to track the time-evolution in the original Hilbert space. {Using this formalism, we developed a low-frequency perturbation theory, which connects naturally with the high-frequency expansions of the Floquet dynamic.}

While we reproduced some results already reported in the literature, for example in two-level systems, our approach allowed us to clarify and extend the range of applicability of these results. Additionally, using the formalism in this paper one can readily obtain the full-cycle dynamics not usually accessible in traditional approaches. For example, the usual treatment of the Landau-Zener model assumes an infinite duration for the transition. However, the actual transition takes a finite time~\cite{Ber90a,SunZhoXia16a}. In experimentally relevant situations, the drive may take a comparable time as the time needed for the transition. In these cases, the Floquet perturbation theory is useful and provides a more detailed description of the dynamics within the drive cycle.

In the low-frequency limit, we obtained a systematic and compact derivation of the adiabatic perturbation theory. Moreover, in this formalism, the occurrence of quasienergy degeneracies that cause diabatic deviations via Rabi oscillations is easily and accurately captured using a degenerate low-frequency Floquet perturbation theory. We saw that for typical, smooth drive protocols, such as sinusoidal, the approach to the adiabatic limit is exponential since the matrix element leading to Rabi oscillations vanishes (super-)exponentially in $1/\Omega$. However, for drive protocols with sharp features, such as step-wise, this matrix element may approach an asymptotic value, thus invalidating the adiabatic approximation. In such scenarios, the Floquet perturbation theory is essential for describing the correct dynamics of the system at low frequencies.

We briefly mention some interesting problems for future application of our work. On the technical side, it would be useful to improve the Floquet perturbation theory for the expansion of the Floquet modes by ensuring the unitarity of the micromotion operator, perhaps along the lines of the Floquet-Magnus expansion~\cite{ManCha11a,EckAni15a}. This would, for example, improve the accuracy of the calculated transition probabilities in the driven Landau-Zener model and avoid divergence for slow drive. Other interesting problems for which the Floquet perturbation theory might be useful are the fate and structure of Floquet topological~\cite{RodSer17a} and Floquet many-body localized~\cite{AbaDe-Huv16a,KheLazMoe16a} phases at low frequencies.

\ack
This work was supported in part 
by 
the National Science Foundation CAREER award DMR-1350663, the US-Israel Binational Science Foundation under grant No. 2014245, and the College of Arts and Sciences at Indiana University (B.S. and M.R.V.), 
as well as the Indiana University REU program through NSF grant PHY-1460882 (M.L.). 
B.S. thanks the hospitality of Aspen Center for Physics, supported by NSF grant PHY-1607611, where parts of this work were performed.

\appendix

\section{Floquet Theory}\label{app:Floquet}

In this Appendix, we provide more detail for the Floquet theory in the extended Floquet Hilbert space.

\subsection{Periodic Hamiltonians}
For a periodic Hamiltonian $\hat H(t)=\hat H(t+T)$, the full-period evolution operator, $\hat U(T)=\Texp\left[-i\int^T_0 \hat H(s)\dd s\right]$, is unitary and can be written as $\exp(-i T \hat H_F)$ for some Hermitian operator $\hat H_F$. The eigenvalues of $\hat U(T)$ are phases $e^{-i\epsilon_\alpha T}$ with eigenstates $\ket{\phi_\alpha}$,
\begin{eqnarray}
\hat U(T) |\phi_\alpha\rangle = e^{-i\epsilon_\alpha T} |\phi_\alpha\rangle. 
\end{eqnarray}
Starting with a state $\ket{\psi_\alpha(0)}=\ket{\phi_\alpha}$, we have $\ket{\psi_\alpha(T)} = \hat U(T)\ket{\phi_\alpha} = e^{-i\epsilon_\alpha T}\ket{\phi_\alpha}$. Thus, defining $\ket{\phi_\alpha(t)} \equiv e^{i\epsilon_\alpha t}\ket{\psi_\alpha(t)}$, we find $\ket{\phi_\alpha(T)} = \ket{\phi_\alpha}\equiv \ket{\phi_\alpha(0)}$, i.e. $\ket{\phi_\alpha(t)}$ are periodic and $\ket{\psi_\alpha(t)} = e^{-i\epsilon_\alpha t}\ket{\phi_\alpha(t)}$. This is the Floquet theorem.

Since $\hat U(t)$ is unitary, it follows immediately that $\epsilon_\alpha\in\mathbb{R}$ and $\{|\phi_\alpha(0)\rangle\}$ is an orthonormal basis for $\mathscr{H}$. Therefore, the $|\phi_\alpha(t)\rangle = e^{i\epsilon_\alpha t}\hat U(t)|\phi_\alpha(0)\rangle$ also form an orthonormal basis for $\mathscr{H}$. Since the quasienergies are modular, defined only through the eigenvalues $e^{-i\epsilon_\alpha T}$ of $\hat U(T)$, we may restrict them to be in the first \emph{Floquet zone}, $\epsilon_\alpha \in [-\Omega/2,\Omega/2]$.
Using this structure, the evolution operator is completely defined by its action $\hat U(t)|\phi_\alpha(0)\rangle = e^{-i\epsilon_\alpha t}|\phi_\alpha(t)\rangle$; thus,
\begin{eqnarray}
\hat U(t)
	= \sum_\alpha e^{-i\epsilon_\alpha t}|\phi_\alpha(t)\rangle\langle\phi_\alpha(0)|
	=: \hat \Phi(t)e^{-i t \hat H_F},
\end{eqnarray}
where the sum over $\alpha$ is understood as an integral whenever $\alpha$ is continuous, and
\begin{eqnarray}
\hat \Phi(t) 
	&= \sum_\alpha |\phi_\alpha(t)\rangle\langle\phi_\alpha(0)|,\\
e^{-i t \hat H_F} 
	&= \sum_\alpha e^{-i t \epsilon_\alpha}|\phi_\alpha(0)\rangle\langle\phi_\alpha(0)|.
\end{eqnarray}
Here $\hat \Phi(t)=\hat \Phi(t+T)$ is a unitary periodic operator, with the boundary condition $\hat \Phi(0)=\hat I$, called the \emph{micromotion operator}. It produces the periodic evolution of the Floquet modes, $\ket{\phi_\alpha(t)} = \hat \Phi(t)\ket{\phi_\alpha(0)}$. The {time-independent}, Hermitian operator $\hat H_F$ is called the \emph{Floquet Hamiltonian}.

\subsection{Micromotion}

In the above decomposition, we chose to set the initial time $t_0=0$. This choice is arbitrary; we could choose any other time within a cycle $0\leq t_0 < T$. In general, for $t>t_0$ we have:
\begin{eqnarray}
\hat U(t,t_0)
	&= \Texp\left[-i\int_{t_0}^t \hat H(s)\dd s \right]  \\
	&= \sum_\alpha e^{-i\epsilon_\alpha (t-t_0)}\ket{\phi_\alpha(t)}\bra{\phi_\alpha(t_0)} \\
	&= \hat \Phi(t,t_0)e^{-i(t-t_0) \hat H_F(t_0)} \\
	&= e^{-i(t-t_0) \hat H_F(t)}\hat \Phi(t,t_0),
\end{eqnarray}
where
\begin{eqnarray}
\hat \Phi(t,t_0) 
	&= \sum_\alpha \ket{\phi_\alpha(t)}\bra{\phi_\alpha(t_0)} = \hat \Phi(t)\hat \Phi(t_0)^\dagger, \\
e^{-i s \hat H_F(t_0)} 
	&= \sum_\alpha e^{-is\epsilon_\alpha}\ket{\phi_\alpha(t_0)}\bra{\phi_\alpha(t_0)} \nonumber\\
	&= \hat \Phi(t_0)e^{-is \hat H_F} \hat \Phi(t_0)^\dagger.
\end{eqnarray}
The two-time micromotion operator produces the periodic evolution of Floquet modes, $\hat \Phi(t,t_0)\ket{\phi_\alpha(t_0)}=\ket{\phi_\alpha(t)}$. 

We note that,
\note{since $[\hat H(t) - i\frac{\dd}{\dd t}]\ket{\phi_\alpha(t)}=\epsilon_\alpha\ket{\phi_{\alpha}(t)}$, the time-dependent Floquet Hamiltonian can be resolved as}
\begin{equation}
\note{\hat H(t) - i\frac{\dd }{\dd t} = \sum_\alpha \epsilon_\alpha \ket{\phi_\alpha(t)}\bra{\phi_\alpha(t)} \equiv \hat H_F(t).}
\end{equation}
Moreover, $\hat H_F$ \note{and $\hat H_F(t_0)$} are unitarily equivalent. Especially, the eigenvalues of $\hat H_F(t_0)$ do not depend on the initial time. However, when certain approximate methods are used to find $\hat H_F(t_0)$ it turns out the eigenvalues acquire a spurious dependence on $t_0$.
One way to avoid this problem is by using the decomposition $\hat \Phi(t,t_0)=\hat \Phi(t)\hat \Phi(t_0)^\dagger$ to write
\begin{eqnarray}
\hat U(t,t_0) = \hat \Phi(t) e^{-i(t-t_0) \hat H_F} \hat \Phi(t_0)^\dagger.
\end{eqnarray}
The dependence on $t_0$ is then entirely accounted for in the micromotion operator. 

\subsection{Floquet Hilbert Space}

The structure we have described in the previous section can be formalized in terms of an extended Floquet Hilbert space $\mathscr{F}=\mathscr{H}\otimes\mathscr{I}$, where the auxiliary space $\mathscr{I}$ is the space of bounded periodic function over $[0,T)$. It is spanned by a continuous orthonormal basis $\{|t)\}, 0\leq t<T$,
\begin{eqnarray}
& (t'|t) 
	= T\delta(t-t'),\\
& \int_0^T |t)(t| \frac{\dd t}T 
	= \invbreve I.
\end{eqnarray}
Here, $\invbreve I$ is the identity operator in $\mathscr{I}$. Equivalently, it is also spanned by the orthonormal Fourier basis 
\begin{eqnarray}
|n) = \int_0^T e^{-in\Omega t}|t)\frac{\dd t}T, \quad n\in\mathbb{Z},
\end{eqnarray}
which satisfy
\begin{eqnarray}
& (n|m) = \delta_{nm},\\
& \sum_{n\in\mathbb{Z}}|n)(n| = \invbreve I.
\end{eqnarray}
Using the Poisson summation formula, $\sum_{n\in\mathbb{Z}}e^{in\Omega t}=T\sum_{p\in\mathbb{Z}}\delta(t-pT)$, these relations can be inverted to give
\begin{eqnarray}
|t) = \sum_{n\in\mathbb{Z}} e^{in\Omega t} |n).
\end{eqnarray}
Note that if we extended the range of $t$ to $\mathbb{R}$ periodically by defining $|t+T)=|t)$, then
\begin{eqnarray}
(t'|t)=T\invbreve\delta(t-t') := T\sum_{p\in\mathbb{Z}}\delta(t-t'-pT) = \sum_{n\in\mathbb{Z}}e^{-in\Omega(t-t')}.
\end{eqnarray}

Now, a loop in $\mathscr{H}$ given by the one-parameter family of states $\ket{\phi(t)}$, with $\ket{\phi(T)}=\ket{\phi(0)}$, can be lifted to a loop in $\mathscr{F}$ given by $\kket{\phi_t}:=\ket{\phi(t)}|t)$. Associated with any loop in $\mathscr{F}$ is the \emph{center}
\begin{eqnarray}
\kket{\overline\phi} := \int_0^T \kket{\phi_t}\frac{\dd t}T.
\end{eqnarray}
From the center of a lifted loop in $\mathscr{F}$ one can in turn obtain the corresponding loop in $\mathscr{H}$ by the projection
\begin{eqnarray}
\ket{\phi(t)} = (t\kket{\overline\phi}.
\end{eqnarray}
We also note that $\kket{\overline\phi}= \sum_{n\in\mathbb{Z}} \ket{\phi^{(n)}}|n),$ where the Fourier components
\begin{eqnarray}
\ket{\phi^{(n)}} = \int_0^T e^{{+}in\Omega t}\ket{\phi(t)} \frac{\dd t}T.
\end{eqnarray}

Similarly, a two-parameter periodic family of operators $\hat A(t,t'):\mathscr{H}\to\mathscr{H}$ with $\hat A(t+T,t')=\hat A(t,t'+T)=\hat A(t,t')$ is lifted to an operator $\hhat{A}:\mathscr{F}\to\mathscr{F}$ defined as
\begin{eqnarray}
\hhat{A} = \int_0^T |t)\hat A(t,t')(t'| \frac{\dd t}T\frac{\dd t'}T = \sum_{n,n'\in\mathbb{Z}}|n)\hat A^{(n,n')}(n'|,
\end{eqnarray}
with
\begin{eqnarray}
\hat A^{(n,n')} = \int_0^T e^{in\Omega t} e^{-in'\Omega t'} \hat A(t,t') \frac{\dd t}T\frac{\dd t'}T.
\end{eqnarray}

Two special cases arise when $\hhat A$ is diagonal in either $\mathscr{I}$ or $\mathscr{H}$.
As an example of the former case, $\hat A(t,t')=\hat A(t)\; T\invbreve\delta(t-t')$ and $\hat A^{(n,n')}=\hat A^{(n-n')}=\int_0^T e^{i(n-n')\Omega t}\hat A(t)\dd t /T$; so,
\begin{eqnarray}
\hhat A = \int_0^T |t)\hat A(t)(t| \frac{\dd t}T = \sum_{n,m\in\mathbb{Z}}|n)\hat A^{(n-m)}(m|.
\end{eqnarray}
As a simple case of the latter case, we take $\hhat A$ to be the identity in $\mathscr{H}$,
so that $\hat A(t,t')$ is just a complex-valued function $A(t,t')$ of its arguments. As an important example, we choose $A(t,t') = e^{{+}in\Omega t}\invbreve\delta(t-t') =:\invbreve \mu_n(t,t')$ to find
\begin{eqnarray}
\hhat \mu_n = \int_0^T |t)e^{{+}in\Omega t}(t|\frac{\dd t}T.
\end{eqnarray}
Then,
\begin{eqnarray}
\kket{\phi_n}
	:=\hhat\mu_n\kket{\overline\phi} = \int_0^T e^{{+}in\Omega t}\kket{\phi_t} \frac{\dd t}T. 
\end{eqnarray}
Thus, $\hhat\mu_n$ acting on the center of a loop in $\mathscr{F}$ yields the Fourier component $n$ of the loop.

Another important example is obtained by the choice
$A(t,t') = i\frac{\partial}{\partial t}\invbreve\delta(t-t')$,
for which we have the operator
\begin{eqnarray}
\hhat Z_{\mnote{t}} &= \int_0^T |t) i\frac{\partial}{\partial t}\invbreve\delta(t-t') (t'| \frac{\dd t}T \frac{\dd t'}T 
	= \sum_{n\in\mathbb{Z}} n\Omega |n) (n|,
\end{eqnarray}
Then,
\begin{eqnarray}
\hhat Z_{\mnote{t}}\kket{\phi_{t}} 
	&= \int_0^T [i\frac{\partial}{\partial t'}\delta(t'-t)]\ket{\phi(t)}|t')\frac{\dd t'}T \nonumber\\
	&= [i\frac{\dd}{\dd t}\ket{\phi(t)}]|t) \equiv i\kket{({\dd\phi}/{\dd t})_t},
\end{eqnarray}
where $\kket{({\dd\phi}/{\dd t})_t}$ is defined as the loop $\frac{\dd}{\dd t}\ket{\phi(t)}$ lifted to $\mathscr{F}$. 
In particular,
\begin{eqnarray}
\bbra{\phi'_{t'}}\hhat Z_{\mnote{t}}\kket{\phi_{t}}  = \bra{\phi'(t)}i\frac{\partial}{\partial t}\ket{\phi(t)} T\delta(t-t').
\end{eqnarray}
We also note that,
\begin{eqnarray}
\hhat Z_{\mnote{t}} \kket{\overline\phi} = i\kket{\overline{{\dd\phi}/{\dd t}}}.
\end{eqnarray}

A loop of periodic family of bases $\ket{\phi_\alpha(t)}$ for $\mathscr{H}$ can be lifted to a basis $\kket{\phi_{\alpha t}}$ for $\mathscr{F}$, since it may be easily seen
\begin{eqnarray}
\iinner{\phi_{\alpha t}}{\phi_{\beta t'}} = \delta_{\alpha \beta} T\delta(t-t'), \\
\int_0^T\sum_{\alpha} \kket{\phi_{\alpha t}}\bbra{\phi_{\alpha t}} \frac{\dd t}{T} = \hhat{I}.
\end{eqnarray}
A different basis is obtained by the Fourier transform of the loop in $\mathscr{F}$, i.e. $
 \kket{\phi_{\alpha n}}=\hhat \mu_n \kket{\overline{\phi_\alpha}}$. To see that this is a complete basis for $\mathscr{F}$, first note that
\begin{eqnarray}
\iinner{\phi_{\alpha n}}{\phi_{\beta m}}
	&= \int_0^T e^{{-}in\Omega t}e^{{+}im\Omega t'} \iinner{\phi_{\alpha t}}{\phi_{\beta t'}} \frac{\dd t}T\frac{\dd t'}T \nonumber\\
	&= \delta_{\alpha\beta} \int_0^T e^{i({m-n})\Omega t} \frac{\dd t}T 
	= \delta_{\alpha\beta}\delta_{nm};
\end{eqnarray}
and, second,
\begin{eqnarray}
\sum_{\alpha n}\kket{\phi_{\alpha n}}\bbra{\phi_{\alpha n}}
	&= \int_0^T \sum_{\alpha n} e^{{-}in\Omega(t-t')} \kket{\phi_{\alpha t'}}\bbra{\phi_{\alpha t}} \frac{\dd t}T \frac{\dd t'}T \nonumber\\
	&= \int_0^T \sum_\alpha \kket{\phi_{\alpha t}}\bbra{\phi_{\alpha t}} \frac{\dd t}T 
	= \hhat I.
\end{eqnarray}
For a time-\emph{independent} basis $\ket{\phi_\alpha}$ in $\mathscr{H}$, we obtain $\kket{\phi_{\alpha n}}=\ket{\phi_\alpha}|n)$, which is obviously a basis for $\mathscr{F}$.

Now, we can write the Floquet Schr\"odinger equation in $\mathscr{F}$ as
\begin{eqnarray}
(\hhat H - \hhat Z_{\mnote{t}}) \kket{\overline{\phi_\alpha}} = \epsilon_\alpha\kket{\overline{\phi_\alpha}}.
\end{eqnarray}
Note that
\begin{eqnarray}
{[ \hhat \mu_n, \hhat Z_{\mnote{t}}]} = n\Omega\hhat \mu_n.
\end{eqnarray}
Since $[\hhat H,\hhat\mu_n] = 0$, we see that $\hhat \mu_n$ is a ladder operator for $\hhat H - \hhat Z_{\mnote{t}}$, mapping the solution $\kket{\overline{\phi_\alpha}}$ with quasienergy $\epsilon_\alpha$ to $\kket{\phi_{\alpha n}}$ with quasienergy $\epsilon_\alpha {+} n\Omega$. Thus, the solutions to the Floquet Schr\"odinger equation
\begin{eqnarray}
(\hhat H - \hhat Z_{\mnote{t}})\kket{\phi_{\alpha n}} = \epsilon_{\alpha n} \kket{\phi_{\alpha n}},
\end{eqnarray}
where $\epsilon_{\alpha n} := \epsilon_\alpha {+} n\Omega$ provide a full basis for $\mathscr{F}$.

\section{Low-frequency Floquet perturbation theory of driven Landau-Zener model}\label{app:LZinFAPT}
In this Appendix, we derive analytic expressions for the driven Landau-Zener model in the low-frequency regime. 

The gauge $\Lambda$ in the adiabatic solutions, Eq.~(\ref{eq:LZad}), is
\begin{eqnarray}
\Lambda(\tau)
	&= \frac1\Omega \int_0^\tau [E_+(s)-\epsilon_{+(0)}]ds \nonumber \\
	&= \frac b\Omega\left[E \left(f_\tau\pi/2 | -a^2/b^2\right)- f_\tau E(-a^2/b^2)\right],
\end{eqnarray}
with $E(\rho|x)=\int_0^\rho\sqrt{1-x\sin^2\alpha}\dd\alpha$ an incomplete elliptic integral, and $f_\tau=2\{\tau/\pi\}-1$ a periodic saw-tooth function ($\{x\}$ is the fractional part of $x$). For asymptotic values of $a/b$, we may expand the elliptic functions to find
\begin{eqnarray}
\Lambda(a\ll b) &= \frac{a^2}{8b\Omega}\sin(2\tau) + O(a/b)^3, \\
\Lambda(a\gg b) &\approx A\frac{a}\Omega \sin(2\tau) + O((b/a)\log(b/a)),
\end{eqnarray}
For $a\gg b$, we have substituted $\sgn(f_\tau)[1-|f_\tau|-\cos(f_\tau\pi/2)]) \approx A \sin(2\tau)$, with $A=(\sqrt{2}-1)/2$.

For these asymptotic expressions of the gauge, $\Lambda=\Lambda_0\sin(2\tau)$, we expand
\begin{eqnarray}
e^{2i \Lambda_0\sin(2\tau)} = \sum_{k\in\mathbb{Z}} J_k(2\Lambda_0) e^{2ik\tau},
\end{eqnarray}
in Bessel functions $J_k$. Then, using the integral
\begin{eqnarray}
\int_0^{2\pi} \frac{x\sin\tau e^{i m\tau}}{1+x^2\cos^2\tau}  \dbar\tau = i^m (\sqrt{1+1/x^2}-1/x)^{|m|},
\end{eqnarray}
for odd $m$ and the fact that the integral vanishes for even $m$, we find $z_n=0$ for even $n$ and, for odd $n = 2q+1$,
\begin{eqnarray}
z_{2q+1} 
	&= \frac i2 \sum_{k\in\mathbb{Z}} (-1)^{q+k} J_{2k}(\Lambda_0) y^{|2(q+k)+1|} \\
	&= (-1)^q i \sum_{p=0}^\infty J_{q,p}(\Lambda_0) y^{2p+1}
\end{eqnarray}
where $y=\sqrt{1+(b/a)^2}-(b/a)\in[0,1]$, and $J_{q,p}(x) = \frac12[J_{q-p}(2x) + J_{q+p+1}(2x)]$.

For small $\Omega/b\ll 1$, the smallest shift is obtained when $a/b\ll 1$ to be $q\sim b/\Omega$. In this limit, $y \approx a/2b$. Assuming $a/\Omega \lesssim 1$, we find $\Lambda_0 = a^2/(8b\Omega)\ll 1$. The biggest contribution to $z_{2q+1}$ would not come from the smallest power of $y$, since this is multiplied by $J_q(2\Lambda_0)$, which is suppressed by the factorial $q!$. Instead, it comes from the term with $J_0$, that is
\begin{eqnarray}
z_{2q+1} 
	\approx y^{2q+1} 
	\sim (\Omega/b)^{2b/\Omega},
\end{eqnarray}
as discussed in Sec.~\ref{sec:LZdFAPT}.

\section*{References}

\end{document}